\documentclass[fleqn,usenatbib]{mnras}

\usepackage{newtxtext,newtxmath}

\usepackage[T1]{fontenc}

\DeclareRobustCommand{\VAN}[3]{#2}
\let\VANthebibliography\thebibliography
\def\thebibliography{\DeclareRobustCommand{\VAN}[3]{##3}\VANthebibliography}

\usepackage{graphicx}	
\usepackage{amsmath}	
\usepackage{pgf}

\graphicspath{{./}}

\title[Extreme Pebble Accretion]{Extreme Pebble Accretion in Ringed Protoplanetary Discs}

\author[Cummins, D. P. et al.]{
Daniel P. Cummins,\thanks{E-mail: daniel.cummins17@imperial.ac.uk (DPC)}
James E. Owen
and Richard A. Booth
\\
Astrophysics Group, Department of Physics, Imperial College London, Prince Consort Rd, London, SW7 2AZ, UK
}

\date{Accepted XXX. Received YYY; in original form ZZZ}

\pubyear{2022}

\begin{document}
\label{firstpage}
\pagerange{\pageref{firstpage}--\pageref{lastpage}}
\maketitle

\begin{abstract}
  Axisymmetric dust rings containing tens to hundreds of Earth masses of solids have been observed in protoplanetary discs with (sub-)millimetre imaging. Here, we investigate the growth of a planetary embryo in a massive ($150 M_\oplus$) axisymmetric dust trap through dust and gas hydrodynamics simulations. When accounting for the accretion luminosity of the planetary embryo from pebble accretion, the thermal feedback on the surrounding gas leads to the formation of an anticyclonic vortex. Since the vortex forms at the location of the planet, this has significant consequences for the planet's growth: as dust drifts towards the pressure maximum at the centre of the vortex, which is initially co-located with the planet, a rapid accretion rate is achieved, in a distinct phase of ``vortex-assisted'' pebble accretion. Once the vortex separates from the planet due to interactions with the disc, it accumulates dust, shutting off accretion onto the planet. We find that this rapid accretion, mediated by the vortex, results in a planet containing $\approx 100 M_\oplus$ of solids.  We follow the evolution of the vortex, as well as the efficiency with which dust grains accumulate at its pressure maximum as a function of their size, and investigate the consequences this has for the growth of the planet as well as the morphology of the protoplanetary disc. We speculate that this extreme formation scenario may be the origin of giant planets which are identified to be significantly enhanced in heavy elements.
\end{abstract}




\section{Introduction}
\label{sec:intro}
High spatial resolution observations of young stellar objects have revealed the wealth of substructure present in their protoplanetary discs. Initially proposed to explain the lack of near-infrared excess in the spectral energy distributions of ``transition discs'' \citep[e.g.][]{strom89, skrutskie90, calvet05}, the presence of substructure in the form of inner cavities was confirmed with early (sub-)millimetre observations \citep{brown09, andrews09, hughes09, isella10, andrews11, mathews12, tang12, fukagawa13}. Many discs observed at millimetre wavelengths with the Atacama Large Millimeter/sub-millimeter Array (ALMA) show concentric rings of emission from millimetre-sized dust grains \cite[e.g.][]{hltau15, fedele17, dsharp1, long18}. Under the assumption that the dust is optically thin at these wavelengths, the measured flux traces the dust mass \cite[e.g.][]{hildebrand83}; interpretations of these observations therefore suggest that the dust is concentrated in axisymmetric rings \cite[e.g.][]{dsharp6}. A handful of discs show non-axisymmetric substructure in their millimetre emission. More specifically, dust in these discs appears to be concentrated azimuthally, forming crescent-shaped features. These asymmetries can be small in scale and exist alongside axisymmetric features (e.g. \citealt{dsharp1}: HD 143006, HD 163296 and \citealt{hashimoto21}: DM Tau), while in the most extreme cases all the millimetre emission comes from one large-scale crescent (e.g. \citealt{vdm13}: IRS 48; \citealt{ohashi08, casassus13}: HD 142527; \citealt{perez14}: SAO 206462, SR 21). Understanding the origin of these substructures, in particular if and how they are related to planet formation, is key to understanding the formation and architecture of planetary systems, as well as reconciling our understanding of planet formation with the observed exoplanet population. 

One of the primary confounding factors in interpreting the observations is that they generally probe the dust component, which does not necessarily directly trace the gas that drives the dynamics. \cite{whipple72} demonstrated how drag forces exerted on dust grains by the gas can cause dust to become trapped at a local axisymmetric gas pressure maximum; such substructure has therefore been interpreted as being due to the presence of dust traps, with the traps forming due to local maxima in gas pressure \cite[e.g.][]{pinilla12, dsharp6, rosotti20}. Observational confirmation of this remains elusive though, as the corresponding gas substructures are yet to be identified. The most abundant gas species observable at millimetre wavelengths are optically thick, and those that aren't suffer from low signal-to-noise limitations. Furthermore, conversion of these species' densities to an overall gas density requires intimate knowledge of the ongoing chemistry; it is therefore unknown whether gas pressure maxima capable of trapping dust at the observed locations exist. Fortunately, the MAPS program \citep{Oberg2021} is beginning to explore this observationally \citep{Law21}, and techniques are being developed to make inferences from gas kinematics \citep[e.g.][]{Teague2019}.

A variety of plausible mechanisms have been shown to form axisymmetric pressure maxima, e.g. planet-carved gaps \citep{lin_papaloizou79, rice06, zhu12}, photoevaporation \citep{Clarke01, alexander06, alexander07, owen2019}, magnetohydrodynamical effects \citep{johansen09, uribe11, bai_stone14}, processes which reduce the local viscosity \citep{johansen11, dullemond18} and processes that can naturally increase the local dust density, such as the freeze-out of gaseous species at their respective icelines \citep{cuzzi2004, kretke07, Okuzumi2016, Owen2020}. These can explain the origin of the observed rings, each with varying degrees of success; \cite{andrews20} provides a comprehensive review. Similarly, a number of mechanisms have been proposed to explain the origin of the observed non-axisymmetric substructures. One interpretation of these observations is that as-yet undetected stellar/substellar companions are present in these systems, which drive eccentricity in the discs, and the observed asymmetry arises simply due to the reduced orbital velocity of material at apocentre \citep{ataiee13}. However, the azimuthal density contrast achieved in this process is inconsistent with observations. Companions on eccentric orbits can also generate similar signatures as they carve eccentric cavities in the circumbinary disc, creating a non-axisymmetric density profile at the cavity edge \cite[e.g.][]{ragusa17, poblete19, ragusa20}, though it's difficult to understand how this mechanism can explain the more localised trapping of larger particles seen at longer wavelength observations, e.g. IRS 48 \citep{vdm13, vdm15}. An alternative explanation proposed by \cite{mittal15} is that, since any inherent large-scale azimuthal asymmetry in the disc shifts the barycentre of the system, in sufficiently massive discs this can give rise to an $m=1$ azimuthal mode, causing the gas to form a horseshoe-shaped annulus that concentrates dust grains via drag \citep{zhu_baruteau16, baruteau2016}. 

The most prevalent interpretation, however, is that an anticyclonic vortex forms in the gas, trapping the dust azimuthally \cite[e.g.][]{LyraLin13}. An anticyclonic vortex is befitting as it represents a pressure maximum and is therefore capable of trapping dust in a similar fashion to the axisymmetric pressure maximum described above \cite[e.g.][]{barge95, tanga96, klahr_henning97, godon99, godon00}. Moreover, anticyclonic vortices are stable to the intrinsic shear of the Keplerian flow and therefore have the potential to be long-lived. However, it is uncertain whether they are stable in real, three-dimensional discs, where they are subject to dissipation by viscosity \cite[e.g.][]{meheut12}, in addition to destruction via elliptical instabilities \citep{kerswell02, lesur09, lyra11} and possibly even through the accumulation of dust \cite[e.g.][]{fu14, raettig15, lyra18}.

Vortex formation in protoplanetary discs has been shown to proceed through various mechanisms. \cite{klahr_bodenheimer03} demonstrated that in discs with a radial entropy gradient, azimuthal perturbations are baroclinically unstable and can lead to the formation of vortices.
The most commonly invoked mechanism to form a large-scale vortex in a protoplanetary disc is the Rossby wave instability (RWI). The RWI can be triggered at steep radial density gradients \citep{lovelace99, li00, li01}; such a gradient may be generated at the viscosity transition between regions in which the magnetorotational instability can operate \citep[e.g.][]{varniere_tagger06, lyra09b, lyra_maclow12, regaly12}. Steep radial density gradients are also found at the outer edge of a gap carved in the disc by a giant planet. Indeed, hydrodynamics simulations of protoplanetary discs containing a giant planet have successfully produced large-scale vortices \cite[e.g.][]{li05, masset06, devalborro07, meheut10, lin_papaloizou11, lin12}, and \cite{regaly17} presented a comparison of vortices formed at a gap edge to those formed at a viscosity transition. However, \cite{hammer17} pointed out that the short growth time-scales of the giant planet prescribed in these simulations result in unphysically large perturbations which are capable of setting up the RWI-unstable density gradient at the gap's outer edge. When the giant planet is grown over more realistic time-scales, vortex production is significantly suppressed because viscosity has time to smooth out any steep gradients and the planetary torque itself can reshape the gap edge. Under the most favourable conditions for vortex formation, vortex lifetimes were found to be limited. One may then argue that if these asymmetries are indeed vortices formed through the RWI triggered by giant planets, we must be seeing them just after the planet has formed. However, if a RWI-generated vortex typically survives for $\sim 10^3$ orbits and discs are $\sim 10^5$ orbits old, observing the vortex so soon after its formation is unlikely. It is therefore uncertain whether the non-axisymmetric substructures seen in observations can be explained by this mechanism.

\cite{owen17} proposed an alternative vortex formation process, with vorticity instead sourced by the accretion luminosity of a forming planetary core. As a planetary embryo accretes solid material, the potential energy liberated by accretion locally heats the disc. For a sufficiently high accretion rate, the thermal structure of the disc can be modified beyond the embryo's Hill sphere, making the disc locally baroclinic and therefore unstable to vortex formation. Hydrodynamics simulations of the circumstellar gas, heated in the vicinity of an accreting planetary embryo, showed that a large-scale vortex forms within tens of orbits and survives for hundreds. However, since these simulations did not explicitly include the dust component of the disc, they required the use of a prescription for the pebble accretion rate and hence the accretion luminosity. The aim of this work is therefore to compute the evolution of the dust along with the gas, enabling a self-consistent calculation of the pebble accretion rate onto the embryo. Specifically, we want to understand if and how the vortex formation process differs when the accretion rate is coupled to the local dust surface density, how long the vortex survives for, and whether it is possible to produce non-axisymmetric dust substructure similar to those present in observations. Also of interest is the mass of the planet formed through this process, and whether we can make a prediction of what mass of planet we might suspect to be present in discs showing similar asymmetries.

We first present the theoretical details of this vortex formation process, followed by a description of the numerical implementation, where we motivate our choice of disc parameters. We then present results of dust and gas hydrodynamics simulations, as well as Monte Carlo radiative transfer simulation results, providing a means of comparison to observations. We discuss the outcome of our simulations in terms of the disc's evolution and the planet's growth. Finally, we assess the limitations of our simulations and provide a comparison to previous work.

\section{Physical Mechanism}
In this section we detail the stages of the planet formation scenario we wish to investigate, starting from the trapping of dust in an axisymmetric gas pressure maximum and how this promotes the formation of a planetary embryo. We then outline how the embryo grows by accreting from the surrounding solid material in the dust trap, and how this can lead to the formation of an anticyclonic vortex.

\subsection{Embryo formation}
\label{sec:embform}
Imposing cylindrical radial ($r$) force balance for the gas component of a protoplanetary disc with vertically-integrated density, $\mathit{\Sigma}_\mathrm{g}$, and pressure, $\mathcal{P}$, yields the orbital speed of the gas as \citep[e.g.][]{armitage2010}:
\begin{equation}
  v_{\phi,\mathrm{g}} = v_\mathrm{K}(1 - \eta)^{1/2},
  \label{eqn:gasorbvel}
\end{equation}
where $v_\mathrm{K}$ is the Keplerian orbital speed and
\begin{equation}
  \eta = -\frac{c_\mathrm{s}^2}{v_\mathrm{K}^2} \frac{\partial \log \mathcal{P}}{\partial \log r}
  \label{eqn:etaitosigma}
\end{equation}
is the ratio of the force due to the thermal pressure gradient to the gravitational force on a gas fluid element due to the central star, where $c_\mathrm{s}$ denotes the isothermal sound speed of the gas \citep{adachi76, weidenschilling77}. The acceleration of a dust grain due to this pressure force is negligible, therefore its orbital speed is unaffected directly by pressure. However, this difference in orbital speeds results in a drag force on the dust, causing it to drift towards the pressure maximum where $\eta$ vanishes. A local gas pressure maximum in a protoplanetary disc will therefore trap dust grains, and those grains with a Stokes number ($\mathrm{St}$) of order unity will be trapped most efficiently \citep[e.g.][]{pinilla2012b}. Under typical protoplanetary disc conditions at tens of au, this corresponds to solids a few centimetres in size, or ``pebbles'':
\begin{equation}
  s \sim 2\:\mathrm{cm} \bigg(\frac{\mathit{\Sigma}_\mathrm{g}}{10\:\mathrm{g\:cm}^{-2}}\bigg)\bigg(\frac{\rho_\bullet}{3\:\mathrm{g\:cm}^{-3}}\bigg)^{-1}\mathrm{St},
  \label{eqn:dustsize}
\end{equation}
where $\rho_\bullet$ is the material density of the dust grain.

With dust trapped at some radius and the continual inward drift of dust from the outer disc, the dust-to-gas mass ratio in the trap is likely to significantly exceed the interstellar medium value of 0.01, potentially reaching values close to unity \cite[e.g.][]{pinilla12, yang17}. It is in such regions of high dust-to-gas ratios that planet formation processes are likely to be enhanced. For example, planetesimal formation can be achieved through the streaming instability \citep{youdin05, johansen07, bai10}, which has been shown to work in regions of protoplanetary discs where dust-to-gas ratios are expected to be enhanced \cite[e.g.][]{drazkowska14, auffinger18} including in dust-traps and locations with low pressure gradient \citep[e.g.][]{carrera21}. Numerical simulations of the streaming instability have demonstrated the formation of planetesimals with masses up to $\sim 10^{-4}M_\oplus$ \cite[e.g.][]{johansen15, simon16, schafer17,abod19}, which can subsequently form planetary embryos through a combination of pebble and planetesimal accretion \cite[e.g][]{liu19, voelkel21_1, voelkel21_2}. It is therefore highly likely that a planetary embryo will form in such a dust trap. 

It is also conceivable that high dust-to-gas ratios are conducive to the formation of multiple embryos; in such a scenario the system would exhibit oligarchic growth, with the first embryo to form dominating the accretion process \cite[e.g.][]{eiichiro98}, as the accretion rate increases rapidly with embryo mass. In summary: axisymmetric dust traps therefore provide ideal conditions for the formation of a planetary embryo, and we argue that it is almost inevitable in a sufficiently strong pressure maximum. In the following modelling we therefore simply invoke the formation of an embryo in the peak of the dust trap.

\subsection{Accretion onto the embryo}
\label{sec:embgrow}
An embryo which forms in an axisymmetric dust trap will find itself in a prime position for growth into a planetary core. Since it is dust grains with Stokes numbers of order unity that are most efficiently and rapidly trapped in pressure maxima, the newly formed embryo will be surrounded by a plethora of these pebbles. \cite{ormel_klahr10} showed that the combined effect of the gravitational potential of the embryo and aerodynamic drag from the surrounding gas causes solids with $\mathrm{St} \sim 1$ to efficiently accrete onto the embryo if they pass within its Hill radius -- the process of pebble accretion. This therefore naturally gives rise to a rapid and efficient core growth scenario, as the dust which is most efficiently trapped also experiences the highest accretion rate onto the embryo.

Pebble accretion has been shown to take place in two distinct regimes \citep{lambrechts_johansen12, morbidelli15}. For a sufficiently massive embryo (mass $M_\mathrm{P}$) orbiting a star of mass $M_*$ at an orbital separation $a$, the Hill radius,
\begin{equation}
  R_\mathrm{H} = a\bigg(\frac{M_\mathrm{P}}{3M_*}\bigg)^{1/3},
  \label{eqn:hillrad}
\end{equation}
will exceed the pebble scale height, allowing accretion to take place in a planar manner, while less massive embryos accrete material in a spherical manner from a fraction of the pebble scale height. Approximating the scale height for dust grains of a given Stokes number as $H_\mathrm{d} \sim H_\mathrm{g}\sqrt{\alpha/\mathrm{St}}$ \cite[e.g.][]{youdin_lithwick07}, where $H_\mathrm{g}$ is the gas pressure scale height and $\alpha$ is the Shakura-Sunyaev turbulent viscosity parameter \citep{shakura73}, the transition between these two regimes can therefore be expressed in terms of a planetary mass threshold, given by
\begin{equation}
  M_\mathrm{P} \gtrsim 10^{-3}M_\oplus \bigg(\frac{M_*}{1M_\odot}\bigg) \bigg(\frac{H_\mathrm{g}/r}{0.1}\bigg)^3 \bigg(\frac{\alpha}{10^{-4}}\bigg)^{3/2}.
  \label{eqn:transmass}
\end{equation}
\cite{lambrechts_johansen12} showed that a planetary embryo with mass above this threshold will accrete pebbles at a rate
\begin{equation}
  \dot{M}_\mathrm{peb} \approx 2\mathit{\Omega}_\mathrm{K}(a)R_\mathrm{H}^2\mathit{\Sigma}_\mathrm{d}(a),
  \label{eqn:m_peb_dot}
\end{equation}
where $\mathit{\Sigma}_\mathrm{d}$ is the surface density of $\mathrm{St} = 1$ pebbles and $\mathit{\Omega}_\mathrm{K}$ is the Keplerian orbital frequency. Estimates of dust masses in ringed discs from their millimetre emission range from around $50 M_\oplus$ \cite[e.g.][]{long18, dsharp6} to around $150 M_\oplus$ \cite[e.g.][]{andrews11}. Dust surface densities derived from these observations suggest a typical accretion rate of
\begin{equation}
  \begin{split}
    \dot{M}_\mathrm{peb} \approx & \: 0.25M_\oplus\:\mathrm{kyr}^{-1} \\
    & \times \bigg(\frac{M_\mathrm{P}}{1M_\oplus}\bigg)^{2/3} \bigg(\frac{M_*}{2M_\odot}\bigg)^{-1/6} \bigg(\frac{a}{\mathrm{35\:au}}\bigg)^{1/2} \bigg(\frac{\mathit{\Sigma}_\mathrm{d}}{1\:\mathrm{g\:cm}^{-2}}\bigg).
    \label{eqn:m_peb_dot_num}
  \end{split}
\end{equation}
Should an embryo form, accretion will therefore be rapid, and the available mass in the trap provides the potential for a massive planetary core to form.

\subsection{Vortex formation}
\label{sec:vortform}
The intrinsic axisymmetric nature of smooth Keplerian discs makes them stable to vortex formation. In the absence of viscosity, the rate of change of vortensity, $\boldsymbol{\omega}/\mathit{\Sigma}_\mathrm{g}$, is given by \citep[e.g.][]{PapaloizouLin95}:
\begin{equation}
  \frac{\mathrm{D}}{\mathrm{D}t}
  \bigg(\frac{\boldsymbol{\omega}}{\mathit{\Sigma}_\mathrm{g}}\bigg) = \frac{\boldsymbol{\nabla}\mathit{\Sigma}_\mathrm{g} \times \boldsymbol{\nabla}\mathcal{P}}{\mathit{\Sigma}_\mathrm{g}^3},
  \label{eqn:vortensity}
\end{equation}
where $\boldsymbol{\omega}$ is the vorticity of the gas. In general, unperturbed discs are barotropic, with $\mathcal{P}(\mathit{\Sigma}_\mathrm{g})$, so vortensity remains constant. However, the presence of an accreting planetary embryo offers a scenario in which the disc can become locally baroclinic, such that the surface density and pressure gradients are no longer aligned, and therefore provide a source of vortensity \citep{owen17}.

On the basis that the kinetic energy of the accreted pebbles is converted into the embryo's thermal energy and ultimately released as radiation, the accretion rate calculated in equation~(\ref{eqn:m_peb_dot_num}) suggests an accretion luminosity of
\begin{equation}
  \begin{split}
    L \approx & \: 5 \times 10^{28}\:\mathrm{erg\:s^{-1}} \\
    & \times \bigg(\frac{M_\mathrm{P}}{1M_\oplus}\bigg)^{17/12} \bigg(\frac{M_*}{2M_\odot}\bigg) \bigg(\frac{a}{\mathrm{35\:au}}\bigg)^{1/2} \bigg(\frac{\mathit{\Sigma}_\mathrm{d}}{\mathrm{1\:g\:cm^{-2}}}\bigg).
  \label{eqn:l_accr_num}
  \end{split}
\end{equation}
This is sufficient to create a hotspot around the embryo that is comparable in size to the embryo's Hill sphere. The pressure structure associated with this hotspot causes gas to deviate from the horseshoe streamlines into circulating either side of the embryo, naturally forming an anticyclonic vortex, where the local rotation is in opposition to the sense of rotation around the central star.

Since an anticyclonic vortex contains a pressure maximum at its centre, solids will drift into the vortex and become trapped. This accumulation of dust will occur at a rate dependent on their Stokes number, maximised for $\mathrm{St} \sim 1$. As discussed above, the grains that contribute most to the pebble accretion rate are also those with $\mathrm{St} \sim 1$. The presence of the vortex therefore has the potential to shut off accretion onto the planet as it traps the pebbles, thereby removing the source of vorticity as well as halting the planet's growth. Similarly, once the vortex dissipates, the dust it accumulated will be released and potentially become available for accretion onto the planet again. \cite{owen17} hypothesised a cycle of vortex formation and dissipation, wherein accretion onto the planet shuts off once the vortex has accumulated the majority of the dust, but once the vortex dissipates and releases the trapped dust, accretion onto the planet may resume, thus restarting the cycle. We therefore wish to investigate the outcome of this process, building on the work of \cite{owen17} by including dust and simulating pebble accretion onto the planetary embryo. Explicitly including the dust allows us to couple the accretion rate of solids onto the planet to the local dust surface density and thus follow the evolution of both the planet and the disc in a self-consistent manner. This provides the ability to see accretion cease once the vortex has formed and trapped enough dust, as well as see any further planet growth once this vortex dissipates.

\section{Numerical Methods}
In order to test our vortex formation scenario we run a number of hydrodynamics simulations of a protoplanetary disc, initialised with an axisymmetric dust trap containing a planetary embryo. We calculate the evolution of both the gas and dust components, and the pebble accretion rate onto the embryo. We perform simulations that both include and exclude the thermal feedback on the disc from the accretion onto the planet, in order to isolate its impact on the outcome of planet growth and the resulting disc substructure.

We use a 2D (vertically integrated) disc model in \texttt{FARGO3D} \citep{fargo3d}, making use of the FARGO algorithm \citep{fargo} to calculate the evolution of the gas, and extensions to the code to calculate the evolution of the dust. Details of the dust dynamics implementation and its validation can be found in \cite{rosotti16}; here we provide a brief summary for completeness before motivating our choice of initial conditions for the gas and dust components as well as the properties of the disc and host star for our simulations. We then describe our implementation of pebble accretion and how we account for the accretion luminosity of the planetary embryo.

\subsection{Dust dynamics}
\label{sec:dustdyn}
The dust is treated as a pressure-less fluid representing grains of a fixed size, coupled to the gas via drag forces. The equation of motion for the dust is therefore
\begin{equation}
    \frac{\partial \boldsymbol{v}_\mathrm{d}}{\partial t} + (\boldsymbol{v}_\mathrm{d} \cdot \boldsymbol{\nabla})\boldsymbol{v}_\mathrm{d} = -\frac{\boldsymbol{v}_\mathrm{d} - \boldsymbol{v}_\mathrm{g}}{t_\mathrm{s}} + \boldsymbol{a}_\mathrm{d},
  \label{eqn:dusteom}
\end{equation}
where $\boldsymbol{v}_\mathrm{d}$ is the dust velocity, $\boldsymbol{v}_\mathrm{g}$ is the gas velocity and $t_\mathrm{s}$ is the time-scale over which the relative velocity decays due to drag, hereafter referred to as the stopping time. The term $\boldsymbol{a}_\mathrm{d}$ captures the acceleration due to forces on the dust other than drag (e.g. stellar gravity). Since \texttt{FARGO3D} calculates the gas velocity in an operator-split manner \cite[e.g.][]{zeus2d, fargo3d}, we calculate the dust velocity in a similar way \citep{rosotti16}: in the source step the dust velocity is updated semi-implicitly, according to
\begin{equation}
  \begin{split}
    \boldsymbol{v}_\mathrm{d}(t + \Delta t) = & \: \boldsymbol{v}_\mathrm{d}(t)\mathrm{exp}(-\Delta t/t_\mathrm{s}) + \boldsymbol{a}_\mathrm{g}\Delta t \\
    & + [\boldsymbol{v}_\mathrm{g}(t) + (\boldsymbol{a}_\mathrm{d} - \boldsymbol{a}_\mathrm{g})t_\mathrm{s}][1 - \mathrm{exp}(-\Delta t/t_\mathrm{s})].
  \end{split}
  \label{eqn:dustupdate}
\end{equation}
The gas acceleration, $\boldsymbol{a}_\mathrm{g}$, is approximated via the change in gas velocity over the source step and  $\Delta t$ is the time-step calculated from the various constraints implemented in \texttt{FARGO3D} to satisfy the Courant-Friedrichs-Lewy (CFL) condition. Additional time-step constraints for the dust velocity components, identical to those of the gas, are included to ensure stability. We treat the drag force under the Epstein approximation, i.e. such that the drag force depends linearly on the relative velocity between the gas and dust \citep{epstein24, whipple72}, an approximation satisfied by our disc's setup and the focus on the outer regions. In this initial investigation we neglect the back-reaction on the gas from the drag force; while this may affect the width of the initial dust ring and the lifetime of any vortices that form, at this stage we aim to isolate the effects of the thermal feedback.

Dust diffusion is included through the addition of a diffusive velocity during the transport step, prior to dust advection. The diffusive velocity is defined as $\boldsymbol{v}_\mathrm{D} = \boldsymbol{F}_\mathrm{D}/\mathit{\Sigma}_\mathrm{d}$, where $\boldsymbol{F}_\mathrm{D} = -(\nu/\mathrm{Sc})\mathit{\Sigma}_\mathrm{g}\boldsymbol{\nabla}(\mathit{\Sigma}_\mathrm{d}/\mathit{\Sigma}_\mathrm{g})$, \citep{clarke88}. $\mathrm{Sc}$ is the Schmidt number, which represents the ratio of momentum diffusion to mass diffusion. Throughout this work we set $\mathrm{Sc}=1$.

We set a dust surface density floor of $10^{-9}\:\mathrm{g\:cm}^{-2}$ such that if the dust surface density in any cell falls below this value (which may occur in a planet-induced gap, for example) the density in that cell is set to the floor value at the end of the transport step. In addition to flooring the dust surface density in such a cell, we set both the radial and azimuthal dust velocity components for that cell equal to their respective gas velocity components. Since transport is performed component-wise, this latter step corrects for a situation in which a negative surface density would yield an incorrect upwind direction, before being transported along the next direction.

We use four species of dust in our simulations, each corresponding to a population of fixed size, spherical grains. Each species represents grains of $1\:\upmu\mathrm{m}$, $1\:\mathrm{mm}$, $3.16\:\mathrm{mm}$ and $1\:\mathrm{cm}$ in radius respectively. Our choice of maximum grain size is motivated by the assumption that within a dust trap, dust grows up to the fragmentation barrier and achieves a steady-state distribution of grain sizes given by the MRN distribution \citep{birnstiel11}. The number, $n$, of dust grains within each species is therefore given by $\mathrm{d}n/\mathrm{d}s \propto s^{-3.5}$. An individual dust grain's size is related to its stopping time through equation~(\ref{eqn:dustsize}), since $\mathrm{St} = t_\mathrm{s}\mathit{\Omega}_\mathrm{K}$. Under the Epstein approximation for the drag law, the stopping time can be expressed as
\begin{equation}
    t_\mathrm{s} \approx \frac{\rho_\bullet s}{\mathit{\Sigma}_\mathrm{g} \mathit{\Omega}_\mathrm{K}}.
    \label{eqn:tstop}
\end{equation}
All dust species are assumed to have a material density, $\rho_\bullet$, of 3 $\mathrm{g\:cm}^{-3}$. We do not include evolution of the dust population through fragmentation or coagulation.

\subsection{Stellar and disc parameters}
\label{sec:params}
In testing our vortex formation mechanism we use disc parameters similar to those in which the majority of both small- and large-scale, non-axisymmetric substructures have been observed (e.g. IRS 48, HD 143006, HD 163296). We therefore choose to model our system on that of a Herbig Ae/Be star of mass $M_* = 2M_\odot$ and a disc that extends out to 175 au. Since we wish to remain agnostic to the origin of the axisymmetric pressure maximum in our simulation, we use an initial disc structure that contains an inner cavity rather than letting one result from some process (such as planet-disc interactions or photoevaporation) as part of our simulation. We therefore choose a surface density profile that results in a local pressure maximum in the disc, at which point $\eta$ vanishes, but with a gradient shallow enough to prevent RWI onset since we wish to avoid other vortex formation mechanisms. Our gas surface density profile takes the form
\begin{equation}
  \mathit{\Sigma}_\mathrm{g} = \frac{\mathit{\Sigma}_0}{2} \bigg(\frac{r}{R_0}\bigg)^{-1} \bigg[1 + \mathrm{erf}\bigg(\frac{r-R_1}{\sqrt{2}\sigma}\bigg) + \epsilon\bigg],
  \label{eqn:gasdens}
\end{equation}
which mimics an inner cavity and a typical power-law profile at large radii. $R_0$ is a characteristic radius which we set to 35 au. $\mathit{\Sigma}_0$ is the corresponding value of gas surface density, which is set by our total disc mass $M_\mathrm{disc} = 0.05M_*$ (this mass is for our surface density profile, equation~(\ref{eqn:gasdens}), including the cavity), a choice motivated by typical values for discs around similar mass stars when crudely assuming a global dust-to-gas mass ratio of 0.01 \cite[e.g.][]{andrews11, ansdell16, dsharp6}. We choose $R_1=1.2R_0$ and $\sigma = 0.663R_0$, which results in a surface density maximum at $r \approx 50\:\mathrm{au}$. $\epsilon = 10^{-3}$ prevents the surface density from vanishing at small radii in our initial condition. Since we are concerned with regions of the disc $\gtrsim 10$ au, the gas density is sufficiently low ($\mathit{\Sigma}_\mathrm{g} \sim 5\:\mathrm{g\:cm}^{-2}$) that the molecular mean free path is larger than the size of dust grains, which justifies our use of the Epstein approximation. For these choices, the minimum value of the Toomre $Q$ parameter is $\sim$5, hence we neglect self-gravity. According to the fragmentation-limited model of \cite{birnstiel12}, this gas density profile gives a maximum grain size of approximately 3 cm, justifying our choice of grain sizes used.

The vertical scale height of the disc varies with radial distance from the central star, according to
\begin{equation}
  H_\mathrm{g} = h_0 R_0 \bigg(\frac{r}{R_0}\bigg)^{5/4},
  \label{eqn:gasscaleheight}
\end{equation}
where $h_0 = 0.05$, providing a flared disc consistent with a temperature profile of $T_\mathrm{b} \propto r^{-0.5}$ \citep{dalessio01}. The pressure and density are linked through the locally isothermal equation of state
\begin{equation}
  \mathcal{P} = c_\mathrm{s}^2 \mathit{\Sigma}_\mathrm{g},
  \label{eqn:iso_eos}
\end{equation}
thus the gas pressure is maximised at $r \approx R_0 \approx 35\:\mathrm{au}$.

Perhaps the choice that will make the biggest difference to our simulations is the total mass of solids in the dust ring. In our first set of simulations, to stress-test this model for vortex formation, we choose a dust mass corresponding to the highest dust masses inferred from observations. We use an initial dust mass of $150 M_\oplus$, comparable to the dust masses in the most massive transition discs \cite[e.g.][]{andrews11}, noting that the simulations presented here are at the extreme end of the expected outcome of planet formation by pebble accretion in a dust ring. In forthcoming work we will vary the initial dust mass as well as disc and star properties, mapping out the outcomes of planet formation in such a dust ring. Higher dust masses will naturally produce initially higher accretion rates (equation~\ref{eqn:m_peb_dot}) and hence stronger thermal feedback.

To determine our initial dust surface density distribution we numerically solve the steady-state continuity equation for an axisymmetric disc,
\begin{equation}
  \frac{\partial \mathit{\Sigma}_\mathrm{d}}{\partial t} = \frac{1}{r}\frac{\partial (r F_\mathrm{d})}{\partial r} = 0,
  \label{eqn:continuity}
\end{equation}
with the radial dust flux, $F_\mathrm{d}$, given by sum of the advective and diffusive components \citep{clarke88}:
\begin{equation}
  F_\mathrm{d} = \mathit{\Sigma}_\mathrm{d}v_{r,\mathrm{d}} - \nu\mathit{\Sigma}_\mathrm{g}\frac{\partial}{\partial r}\bigg(\frac{\mathit{\Sigma}_\mathrm{d}}{\mathit{\Sigma}_\mathrm{g}}\bigg).
  \label{dustflux}
\end{equation}
We define the kinematic viscosity, $\nu$, via the alpha viscosity prescription \citep{shakura73},
\begin{equation}
  \nu = \alpha c_\mathrm{s} H_\mathrm{g},
  \label{eqn:viscosity}
\end{equation}
with a constant $\alpha = 10^{-4}$, and we calculate the radial velocity of the dust according to
\begin{equation}
  v_{r,\mathrm{d}} = \frac{v_{r,\mathrm{g}}\mathrm{St}^{-1} - \eta v_\mathrm{K}}{\mathrm{St} + \mathrm{St}^{-1}},
\end{equation}
as derived by \cite{takeuchi02}. The radial velocity of the gas, $v_{r,\mathrm{g}}$, is given by the speed of viscous accretion, calculated by \cite{lin_papaloizou86}. We verify our solution using \texttt{FARGO3D}, letting the dust evolve from $\mathit{\Sigma}_\mathrm{d}(r) = 0.01\mathit{\Sigma}_\mathrm{g}(r)$ to a steady state.

Our simulation domain extends radially from $0.54R_0$ to $5.0R_0$. We varied the boundary locations to ensure that our area of interest, and thus our results, were not affected by our choice of domain size. We employ an open boundary condition at the inner disc edge, allowing material to flow out of the computational domain towards the host star. At the outer boundary we mirror the dust and gas radial velocities. Outside of the computational domain we set the dust densities to the floor value since all the dust is located in the pressure trap, while the gas density is kept at the values defined by equation~(\ref{eqn:gasdens}). The active domain contains wave-killing zones near the boundaries in order for any artefacts arising from the use of a finite domain to be suppressed. The inner (-) and outer (+) wave-killing zones extend from their respective boundaries to $r_\mathrm{damp}^\pm = r_\mathrm{bound}^\pm 1.15^{\mp2/3}$. The damping is implemented as described in \cite{devalborro06}, with a damping time-scale of 1.5 orbital periods at the respective boundary radii.

In order to resolve the Hill sphere of the planetary embryo at its initial mass we use variable radial cell spacing, with the highest resolution at the location of the planet. Specifically, we use a radial cell spacing given by $\Delta r \propto r[1-\mathcal{G}(r-a)]$, where $a$ is the semi-major axis of the embryo and $\mathcal{G}(r)$ is a Gaussian function with standard deviation $0.1 a$, which connects logarithmically-spaced cells with a narrow region of increased resolution around the embryo's location \citep[cf.][]{Schulik2020}. While this results in grid cells having a variable aspect ratio, it provides a suitable balance between resolution and numerical expense. Our minimum resolution is set by the requirement of resolving the Hill sphere of the embryo at its initial mass with at least eight cells. We therefore use 1920 cells in the radial direction, 660 of which cover $r \in [0.8, 1.2] a$, and 7680 uniformly-spaced cells in the azimuthal direction, which we found to be a satisfactory compromise between convergence and computational expense (see our resolution study in \textsection\ref{sect:resstudy}).

To isolate the physics in this initial work, the planet is kept on a fixed orbit, i.e. we do not let it migrate radially through the disc. The potential of the planet is smoothed over a distance $\epsilon = 0.35 R_\mathrm{H}$ so as to avoid the singularity at the planet location. The smoothed potential, $\mathit{\Phi}_\mathrm{P}$, takes the form
\begin{equation}
  \mathit{\Phi}_\mathrm{P}(d) = -\frac{GM_\mathrm{P}}{(d^p+\epsilon^p)^{1/p}},
  \label{eqn:potsmooth}
\end{equation}
where we set $p=4$, which allows us to accurately capture the form of the potential at distances down to the Hill radius.

\subsection{Pebble accretion and thermal feedback}
\label{sec:pebaccr}
The goal of our accretion routine is that the physics of pebble accretion is resolved in our simulation. Specifically, we want our simulations to supply the planet's Hill sphere with dust at the appropriate rate set by the physical properties of both the planet and the disc. However, since we clearly cannot resolve accretion directly onto the point-mass planet using a fluid treatment of dust, we must remove dust from the planet's vicinity to prevent unphysically large dust surface densities inside the Hill sphere. Thus, by removing dust from the planet's Hill sphere at a rate appropriate for the physics of pebble accretion, the surface density in the Hill sphere will reach a steady balance between the physically derived supply rate from our simulations and our sensibly chosen removal rate. Provided our removal rate is sensibly chosen (and dependent on the dust surface density in the Hill sphere), if we are resolving the supply rate, the removal rate and hence accretion rate onto the planet will then be independent of the actual form of the removal rate; all that will happen is the surface density in the Hill sphere will moderate itself to match the supply rate provided by the physics of the simulation.

We begin with a planetary embryo of mass $0.1M_\oplus$, positioned at the centre of the dust trap ($a \approx 35$ au), since this is the most likely site of embryo formation. This choice of initial mass is a compromise between being low enough that it is sufficiently close to the mass threshold for planar pebble accretion calculated in equation~(\ref{eqn:transmass}), and high enough that resolving its Hill sphere is feasible. We experimented with different initial masses, finding that this choice does not qualitatively change the scenario our simulations produce, but merely introduces a time offset.

\begin{figure}
  \centering
  \includegraphics{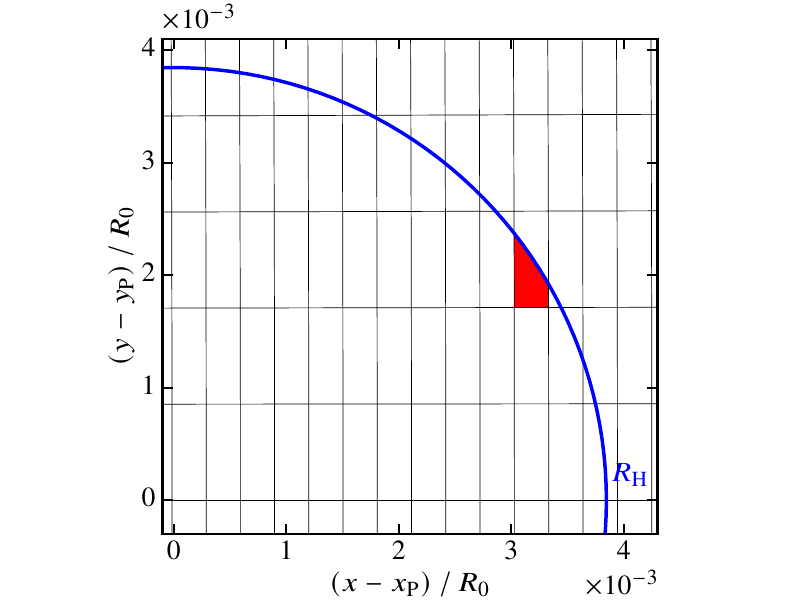}
  \caption{A zoom-in on one quadrant of the planet's Hill sphere, with the coordinate system centred on the location of the planet ($x_\mathrm{P}$, $y_\mathrm{P}$). The black lines mark cell boundaries and the blue line marks the edge of the planet's Hill sphere at its initial mass. The fraction of a cell only partially contained within the Hill sphere is approximated by constructing a polygon within the cell, an example of which is shown in red.}
  \label{fig:cellfraction}
\end{figure}

With an initial mass above the transition mass, the embryo always accretes at a rate roughly given by equation~(\ref{eqn:m_peb_dot}). We therefore use this parametrization as our removal rate of material in the Hill sphere. In practice this accretion rate is calculated numerically by summing the masses in each cell within the Hill sphere; by calculating $\mathit{\Sigma}_\mathrm{d}(a)$ as an average over the Hill sphere, i.e. $\mathit{\Sigma}_\mathrm{d}(a) = M_\mathrm{H}/(\uppi R_\mathrm{H}^2)$ where $M_\mathrm{H}$ is the dust mass in the Hill sphere, we can write
\begin{equation}
  \dot{M}_\mathrm{peb} = \frac{2}{\uppi}\mathit{\Omega}_\mathrm{K}(a)\mathrm{St}^{2/3} \sum_{d \leq R_\mathrm{H}}\mathit{\Sigma}_{\mathrm{d},ij} A_{ij},
  \label{eqn:m_peb_dot_grid}
\end{equation}
where $d$ is the distance from the planet and $A_{ij}$ is the area of the cell with indices $i, j$, which denote the cell's position. It is important to account for the fact that the Hill radius will enclose fractions of cells, as can be seen in figure~\ref{fig:cellfraction}, in order to provide a smooth change in accretion rate as the Hill sphere increases in radius and thus encloses more cells. To estimate the fraction of a cell that lies partially within the Hill radius, we assume that the grid can locally be approximated as Cartesian and that we can therefore treat the cell as a rectangle. The two intercepts that the Hill sphere makes with the cell boundaries are calculated and used to construct a polygon, whose remaining vertices are the cell corners within the Hill radius. The area of this polygon is ultimately evaluated as the sum of areas of a triangle and at most two rectangles, depending on where the Hill sphere crosses the cell. This simple calculation serves as a good approximation for the area of the cell contained within the Hill sphere, yielding a smooth change in accretion rate as the planet's mass increases and new cells enter its Hill sphere.

\begin{figure}
  \centering
  \includegraphics{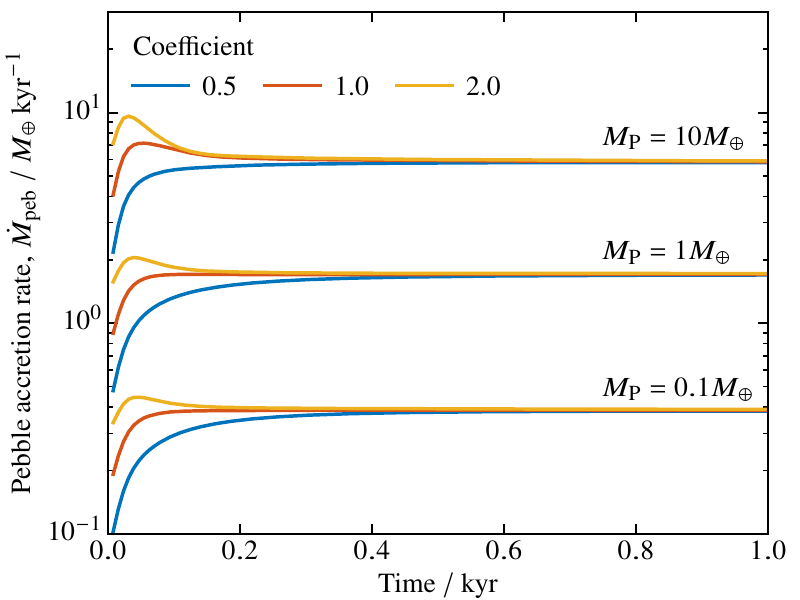}
  \caption{A comparison of the pebble accretion rates for the same initial conditions, but with the accretion rate modified by a constant coefficient. The planet mass is held fixed while dust is removed from its Hill sphere according to the pebble accretion rate, and the test is performed for planets of mass $0.1 M_\oplus$, $1.0 M_\oplus$ and $10 M_\oplus$. Despite the modifying factor, the accretion rates converge to a single value for a given planet mass, as the planet naturally sets the accretion rate by regulating the mass in its Hill sphere.
  }
  \label{fig:ffcomparison}
\end{figure}

The value of $\mathrm{St}$ used in equation~(\ref{eqn:m_peb_dot_grid}) is calculated as an azimuthal average at the semi-major axis of the planet, as equation~(\ref{eqn:m_peb_dot}) implicitly assumes an axisymmetric disc, unperturbed by the presence of the planet. As a means of verifying that we are actually resolving pebble accretion in our simulation, we performed a test wherein we measured the pebble accretion rate onto the planet, held at a fixed mass, and varied the pebble accretion rate as expressed in equation~(\ref{eqn:m_peb_dot}) by factors of 0.5 and 2. The planet is held at a fixed mass because any initial differences in the pebble accretion rate will lead to differences in planet mass and therefore the subsequent accretion rate, making it difficult to distinguish between effects from modifying the accretion rate and effects from differences in planet mass. The results of this test are shown in figure~\ref{fig:ffcomparison}. If the rate of removal of dust from the Hill sphere is controlling the accretion rate we would expect these modifications to linearly change the accretion rates. However, if our simulations are correctly resolving pebble accretion, these modifications should instead have no impact on the actual accretion rate as described above. What these tests show is that the effect of the order-unity modification is simply an initial time offset in the accretion rate, due to the dust in the Hill sphere being unperturbed by the embryo at the start of the simulation. Once the surface density in the vicinity of the planet has reached a pseudo steady-state between supply and removal from the Hill sphere, which takes $\approx 4$ orbits, our accretion rate is insensitive to the exact choice of the mass removal scheme. These tests demonstrate that the accretion rate naturally adjusts itself to the appropriate value as the planet regulates the mass in its Hill sphere, and hence we are resolving pebble accretion as desired. This test was performed for planets of mass $0.1 M_\oplus$, $1 M_\oplus$ and $10 M_\oplus$, demonstrating that the result holds down to our initial mass.

The planet's mass is updated according to $\Delta M_\mathrm{P} = \dot{M}_\mathrm{peb} \Delta t$. We impose an additional time-step constraint such that no more than 10\% of the dust mass in the Hill sphere is removed in a single time-step. The amount of mass $\Delta M_\mathrm{P}$ is removed from the planet's Hill sphere, with the mass in each cell reduced in proportion to the mass there at that instant.

We calculate the accretion luminosity resulting from pebble accretion as the rate of change potential energy lost by the accreted pebbles, i.e.
\begin{equation}
  L = \frac{GM_\mathrm{P}\dot{M}_\mathrm{peb}}{R_\mathrm{P}},
  \label{eqn:L_accr}
\end{equation}
where $R_\mathrm{P}$ is the radius of the planet, which is calculated as a function of its mass following the mass-radius relationship of \cite{fortney07}, assuming an ice mass fraction of 0.5. While this choice of ice mass fraction is motivated purely by the planet's orbital separation of 35 au, it allows us to veer on the conservative side of parameter space since a rocky core would yield a higher accretion luminosity.

The accretion luminosity is used to prescribe the radius of the hotspot around the planet. We work under the assumption that the disc is optically thin to infrared radiation in the vicinity of the planet, as is expected in the outer regions of the disc (and confirmed by our optical depth estimates over the Hill sphere from our simulation, which are of order $\sim 10^{-3}-10^{-1}$), and that the change in luminosity of the planet is instantaneously translated into a change in temperature. The distance at which the blackbody temperature solely due to the planet's radiation would equal the background disc temperature is
\begin{equation}
  R_\mathrm{L} = \sqrt{\frac{L}{16\uppi\sigma T_\mathrm{b}^4(a)}},
  \label{eqn:lumrad}
\end{equation}
where $\sigma$ is the Stefan-Boltzmann constant. The temperature of the disc is therefore approximately given by \citep{owen17}:
\begin{equation}
  T = T_\mathrm{b}\bigg(1 + \frac{R_\mathrm{L}^2}{d^2+s^2}\bigg)^{1/4},
  \label{eqn:disctemp}
\end{equation}
where $T_\mathrm{b}$ is the background disc temperature due to the central star, defined in equation~(\ref{eqn:gasscaleheight}), and $d$ is the distance from the planet. $s$ is a smoothing parameter which we set to be $0.65 R_{\mathrm{H},0}$, where $R_{\mathrm{H},0}$ is the Hill radius of the planet at its initial mass. Since we employ a locally isothermal equation of state, the temperature increase due to the hotspot is added by re-writing the sound speed during the pressure calculation step.

\begin{figure*}
  \centering
  \includegraphics{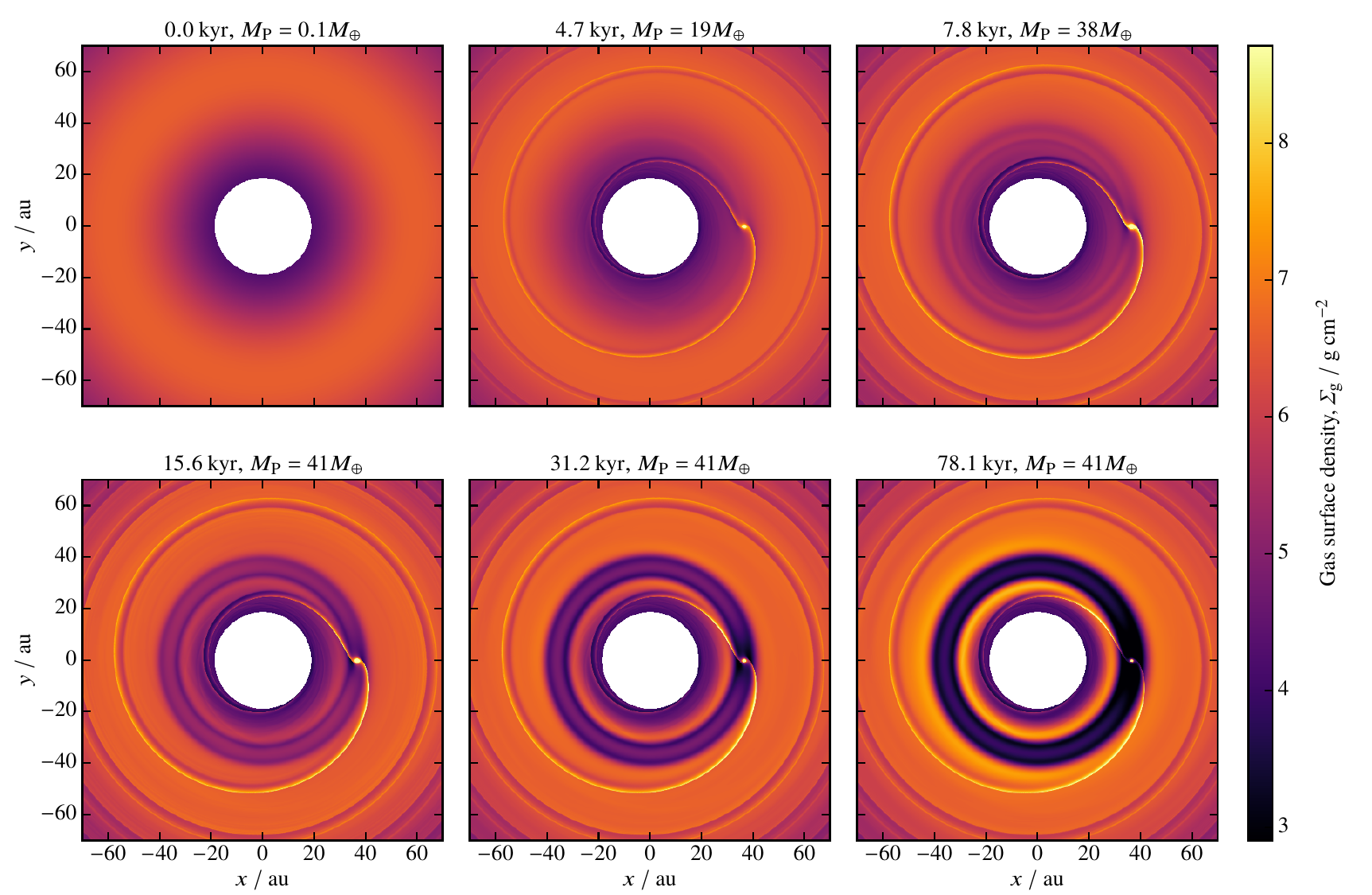}
  \caption{Gas surface density snapshots for the simulation in which the planetary embryo undergoes pebble accretion without heating the disc. The disc is viewed in a frame rotating with the planet, which therefore remains fixed at at (35, 0). Above each panel is the time of the snapshot since the beginning of the simulation and the mass that the planet has grown to. As the planet grows it creates a shallow gap in the disc which inhibits further inward drift of dust, preventing further accretion. Dust within the co-orbital region is confined to horseshoe orbits, which halts the growth of the planet.}
  \label{fig:surfdens_nh}
\end{figure*}

\section{Results}
\subsection{Hydrodynamics simulations}
\subsubsection{No planet test}
In order to verify that our initial gas density distribution is stable against vortex formation via the RWI or otherwise, we first let the disc evolve without the planetary embryo present. Over 150 kyr (equivalent to $\sim1000$ orbits at 35 au) the gas pressure maximum slowly reduces due to viscosity and the dust trap widens. The dust remains trapped in the axisymmetric pressure maximum, which gradually moves towards the star due to viscous accretion. No substructure forms in the gas or dust, confirming that the density gradient maintaining the dust trap is stable to vortex formation.

\subsubsection{No accretion luminosity test}
Since we wish to isolate the effect of the accreting embryo's thermal feedback on the disc, we performed a simulation with the accretion luminosity artificially kept at zero while the embryo accretes pebbles. Figure~\ref{fig:surfdens_nh} shows snapshots of the gas surface density from this simulation. As the planetary embryo accretes the solid material and increases in mass, the strength of its gravitational interaction with the disc increases, launching spiral density waves and carving a gap in the disc. The planet eventually reaches $41 M_\oplus$, by which point it has carved a gap deep enough to prevent further drift of dust into its orbit. Note that this mass is significantly higher than the standard pebble isolation mass that occurs for planets forming in discs without pressure traps. 

After $\sim 35$ kyr the gap edges become unstable to the RWI, which leads to the formation of vortices at either gap edge, which then dissipate fairly quickly. The result of this can be seen in the final panel in figure~\ref{fig:surfdens_nh}, where the gas surface density immediately outside of the gap is slightly enhanced in the shape of an extended crescent. Dust that was present in the co-orbital region but was not accreted onto the embryo accumulates close to the classical L$_5$ Lagrange point, though this does not occur at the opposite Lagrange point, L$_4$. The location of the L$_5$ trap varies between dust species, a physical process we discuss in \textsection\ref{sec:discuss}.

\subsubsection{Impact of accretion luminosity}
\label{sec:mainresult}
\begin{figure*}
  \centering
  \includegraphics{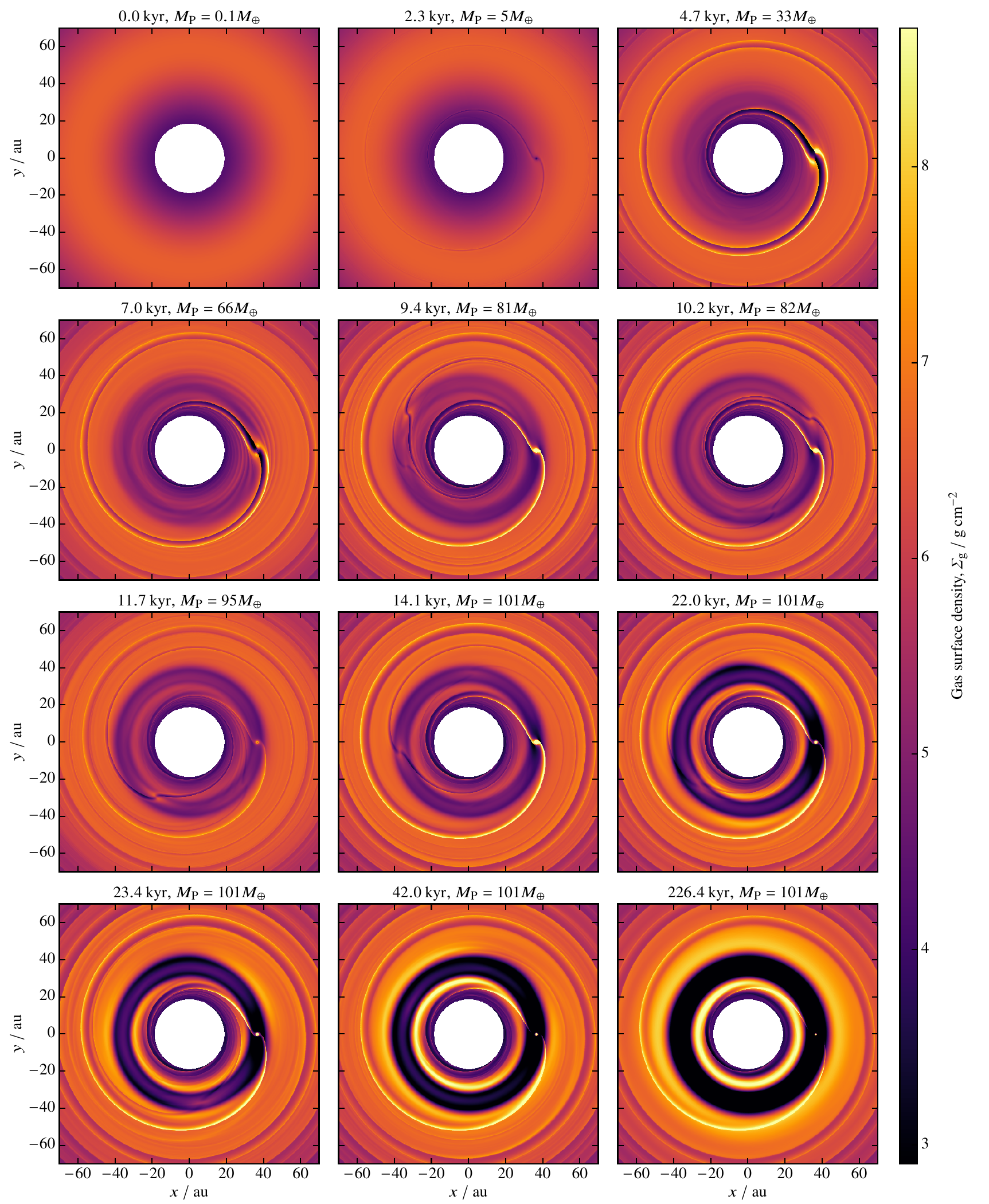}
  \caption{Gas surface density snapshots showing the formation of vortices as the planetary embryo accretes pebbles and heats the surrounding gas. Vortex formation can be seen at the planet location via the enhancement in surface density, and vortices can be seen to have separated from the planet after 7 kyr, again identifiable by their surface density enhancements as well as the spiral shocks they generate. The vortices traverse horseshoe orbits due to the planet's significant mass, and the corotation torque on the vortex from the planet results in material entering the Hill sphere, causing bursts of accretion at $\approx$10.5 kyr and $\approx$12 kyr.}
  \label{fig:surfdens_h}
\end{figure*}

\begin{figure*}
  \centering
  \includegraphics{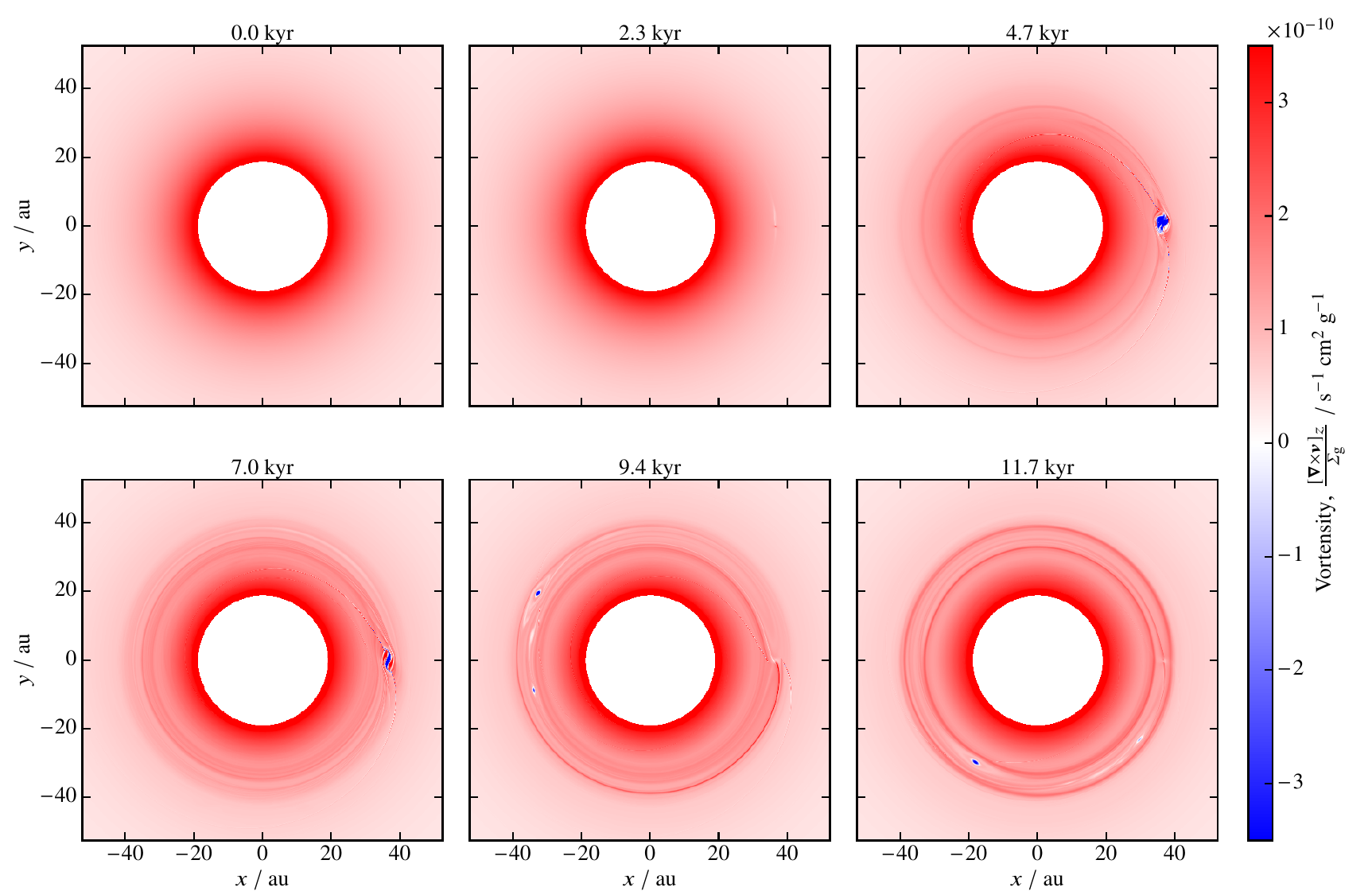}
  \caption{Vortensity snapshots for the simulation including the accretion luminosity of the planet. A region of negative vortensity, indicating the anticyclonic nature of the vortex, can be seen to form at the location of the planet before traversing horseshoe orbits. Two smaller, weaker vortices can be identified from the snapshot at 9.4 kyr onward.}
  \label{fig:vortens_h}
\end{figure*}

With the behaviour of the system due purely to the gravity of the planet now established, we present the results of the simulation where we account for the accretion luminosity of the embryo. Figure~\ref{fig:surfdens_h} shows the evolution of the gas surface density in the disc at a series of snapshots over the $\approx 230$ kyr ($\sim 1600$ orbits) simulation, as well as the mass that the embryo has grown to at each snapshot. As the embryo accretes and the energy released heats the surrounding gas, waves are launched from the hotspot around the embryo. By 4 kyr, an enhancement in gas density in the vicinity of the embryo can be seen, indicating the initial stages of vortex formation are underway. This can be confirmed by inspecting the vortensity, which is shown in figure~\ref{fig:vortens_h}, where a region of negative vortensity is associated with this enhanced surface density. By 7 kyr an anticyclonic vortex has formed at the location of the planet. The vortex's interaction with the disc causes a slight radial migration and subsequent separation of the vortex from the planet's location. Two smaller, weaker vortices can also be seen (e.g. at 9.4 kyr in figure~\ref{fig:vortens_h}), which separate from the planet shortly after the first, largest vortex. The planet becomes massive enough that it carves a deep gap and confines the vortices to horseshoe orbits, before they merge into a single vortex. The resulting vortex survives for $\sim 75$ kyr.

The inner and outer edges of the gap carved by the planet also become unstable to the RWI for a short period of time. At the outer edge, this results in the growth of the $m=3$ mode and the subsequent formation of three vortices, at around 22 kyr, while at the inner edge a single vortex forms. By $\sim 150$ kyr these vortices have dissipated, leaving the disc in its final, stable state, shown in the final panel.

\begin{figure*}
  \centering
  \includegraphics{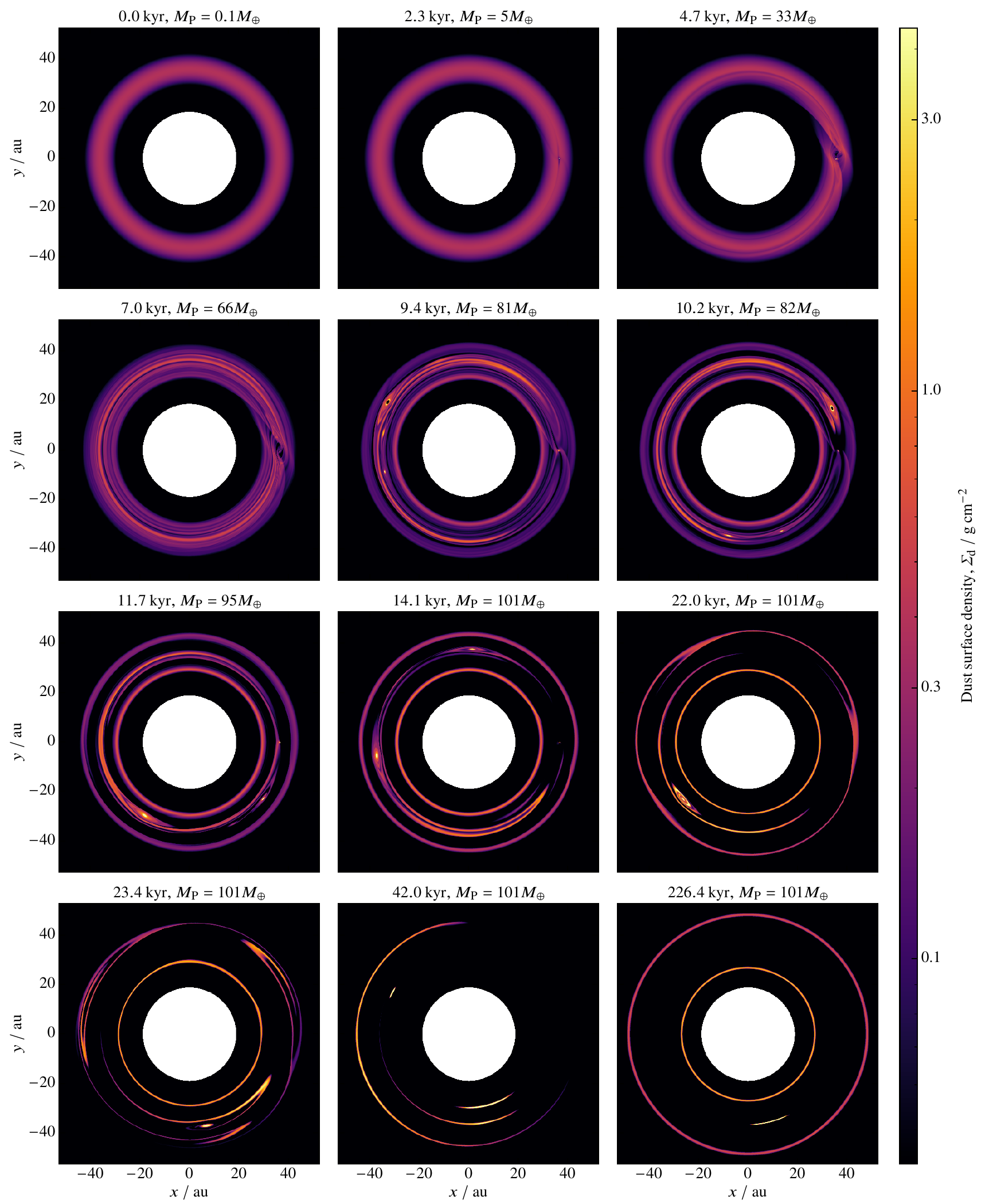}
  \caption{Surface density of 1 mm dust for the simulation including the accretion luminosity of the planet. Initially, the shock through the vortex reduces the pressure maximum at its centre, which prevents dust from directly accumulating there. Once the vortensity has reduced, however, dust can drift into the centre. Dust from the initial ring remains trapped in the pressure maxima either side of planet-carved gap, as well as in the vortices. Once the vortices have dissipated, dust resides close to the L$_5$ Lagrange point.}
  \label{fig:dustdens2_h}
\end{figure*}

Figure~\ref{fig:dustdens2_h} shows the surface density of the 1 mm dust grains at the same snapshots in time as in figure~\ref{fig:surfdens_h}. A high surface density region can be associated with the vortex structures observed in the gas, confirming the ability of the anticyclonic vortices to trap dust. As the planet grows rapidly and carves a gap in the pressure maximum, some dust remains trapped either side of its orbit, unable to drift into the vicinity of the planet to be accreted, hence two axisymmetric rings form either side of the planet's orbit. As the vortex begins to dissipate, the dust it releases back into the co-orbital region begins to accumulate around L$_5$. The snapshots between 22 and 42 kyr show how the vortex interacts with the dust that settles around L$_5$: the vortex attracts this dust to its pressure maximum, but the dust does not have enough time to become trapped in the vortex, and is instead dragged along by it, forming the thin, azimuthally-extended features that can be seen in these later snapshots.

Once the gap edges become RWI-unstable, the axisymmetric dust rings are affected by the vortices that form. The snapshot at 22 kyr shows the dust beginning to concentrate into the $m=3$ vortex pattern in the outer ring. This results in clear asymmetries in the 1 $\upmu$m and 1 mm dust, but not in the 3 mm and 1 cm dust, indicating that these vortices are weak (a result confirmed by inspection of the vortensity) and so only the dust most strongly coupled to the gas shows asymmetries. Once all the vortices have dissipated, dust settles at L$_5$, as was found in the simulation without the thermal feedback. The final state of the disc is therefore two dust rings and dust at L$_5$, where the masses remaining in these three distinct substructures are 13.3, 20.9 and 15.1$M_\oplus$ respectively, summed across all four dust species.

\begin{figure*}
  \centering
  \includegraphics{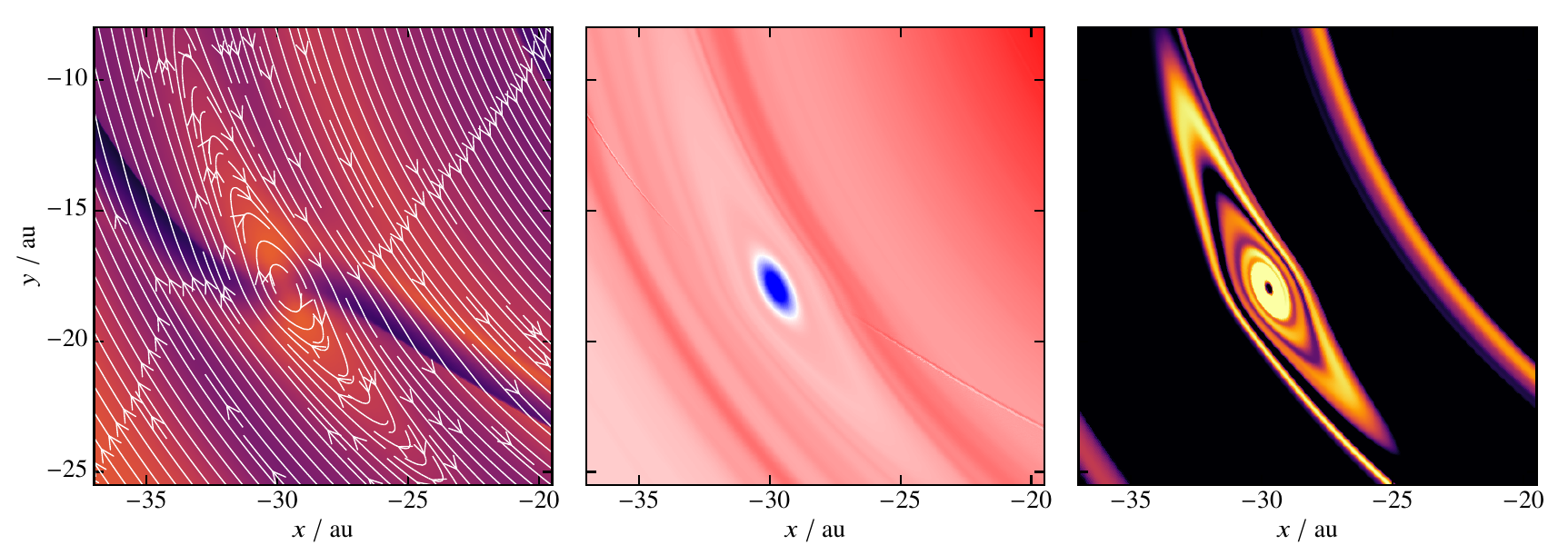}
  \caption{Zoom-in on the vortex at $\approx 12$ kyr, showing the gas surface density with gas streamlines (left), vortensity (centre) and 3 mm dust surface density (right). The colour scales are the same as in figures~\ref{fig:surfdens_h}, ~\ref{fig:vortens_h} and ~\ref{fig:dustdens2_h} respectively. The radial extent of the vortex is approximately 2.5 au, corresponding to $4/3 H_\mathrm{g}$.}
  \label{fig:vortex_figs}
\end{figure*}

Figure~\ref{fig:vortex_figs} shows the gas surface density, gas velocity streamlines, vortensity and dust surface density at the location of the vortex $\approx 12$ kyr into the simulation. The streamlines confirm the anticyclonic nature of the vortex, with the sense of local rotation in opposition to the global flow around the star, and the gas surface density shows that it is launching spiral density waves. The closed streamlines around the vortex span $\approx 2.5$ au in radial extent, equivalent to $\approx 4/3 H_\mathrm{g}$. If the radial extent of a vortex exceeds $2/3 H_\mathrm{g}$ from its centre, the supersonic velocity shear in a Keplerian profile will prevent the vortex from expanding \citep{adams_watkins95}, thus our measured width indicates that the vortex has grown to its largest possible radial extent.

\begin{figure}
  \centering
  \includegraphics{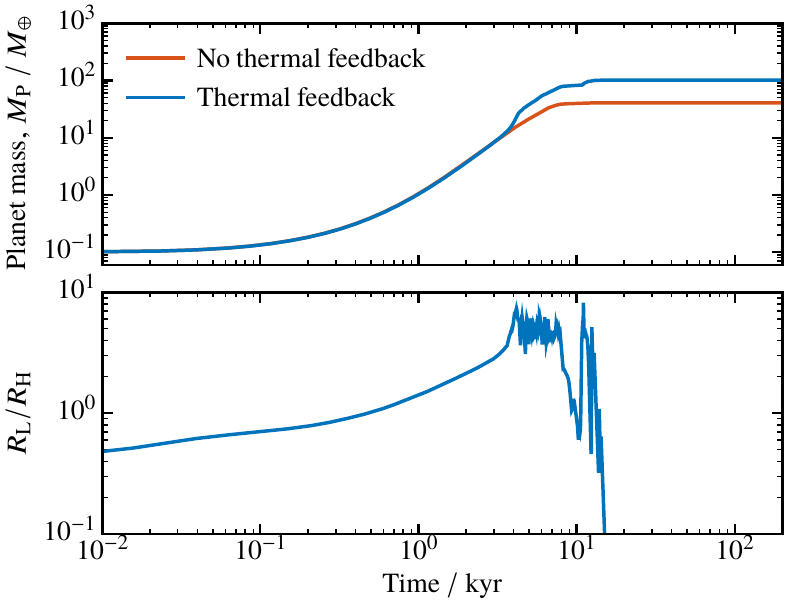}
  \caption{The top panel shows the mass of the accreting planet as a function of time for the simulations with and without the thermal feedback. In both cases the embryo initially grows slowly, but the accretion rate increases as its Hill sphere grows. With the inclusion of the thermal feedback, the high accretion rate leads to vortex formation at the planet location at $\approx 3.5$ kyr. This initially enhances the accretion rate, as the vortex draws dust towards the planet, but once the vortex breaks away from the planet it quickly traps dust, shutting off accretion onto the planet. The bottom panel shows the ratio of the luminosity radius to the Hill radius. Once this ratio exceeds $\sim3$, the thermal feedback is sufficient to generate a vortex that can significantly affect the accretion rate. This ratio drops off once accretion ceases, staying below $\sim 10^{-3}$ after 20 kyr.}
  \label{fig:planetmass}
\end{figure}

A comparison of the planet mass as a function of time for the simulations with and without the thermal feedback is shown in figure~\ref{fig:planetmass}, illustrating the impact of the accretion luminosity on the growth of the planet. Also shown is the ratio of the luminosity radius, $R_\mathrm{L}$, to the Hill radius, $R_\mathrm{H}$. Vortex formation requires the thermal structure of the disc to be disrupted beyond the Hill sphere. The planet growth rates begin to diverge at around 3.5 kyr, corresponding to the time at which the luminosity radius exceeds the Hill radius by a factor $\sim3$. The accretion rate falls off after 7 kyr, once the vortex has separated from the planet location, trapped dust and therefore dramatically reduced the solid material available for accretion onto the planet. There is a second burst in accretion at $\approx 10$ kyr, which results from the corotation torque on the vortex from the planet causing a significant amount of material to enter the planet's Hill sphere. The planet in the simulation with thermal feedback eventually reaches $\approx 100 M_\oplus$, more than twice that achieved in the simulation without, which is a naively surprising result, as we might have expected dust trapping by the vortex to suppress the accretion rate. We discuss the implication of this result in terms of the physics of pebble accretion and vortex formation in \textsection\ref{sec:discuss}.

\begin{figure}
  \centering
  \includegraphics{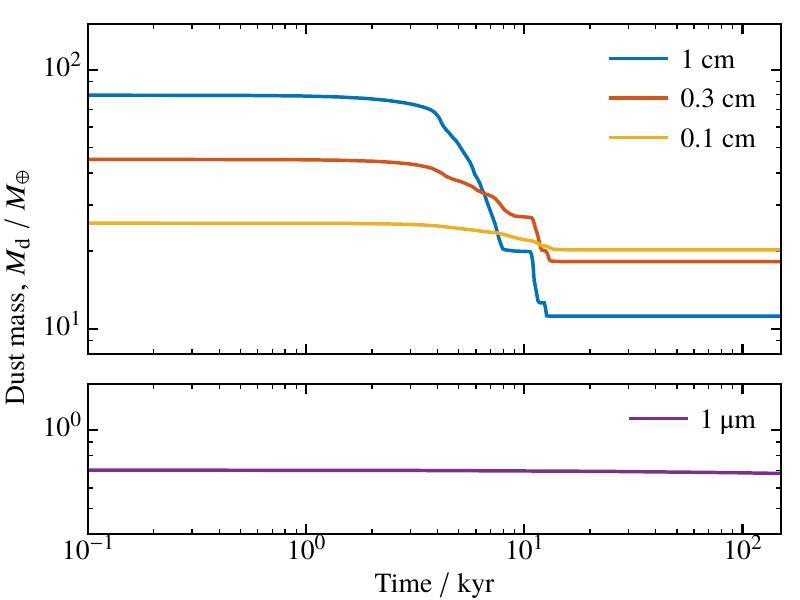}
  \caption{Mass remaining in each dust population as a function of time. The 1 cm population loses the largest fraction of its mass since they have the highest pebble accretion rate, and the dust size distribution is inverted once pebble accretion ceases. The micron sized dust has been shown on a separate panel for clarity. The total dust mass remaining is $50 M_\oplus$, which is trapped in the dust rings either side of the planet's orbit, and in the co-orbital region. The steep drops in mass after 10 kyr are due to the corotation torque between the vortex and planet causing dust to be released in to the planet's Hill sphere.}
  \label{fig:dustmass}
\end{figure}

Figure~\ref{fig:dustmass} shows the mass of solids in the disc over the duration of the simulation. The largest grain size populations, which have Stokes numbers closest to unity, lose the largest fraction of their mass since they have the highest pebble accretion rate onto the planet. Similarly, since they are the most efficiently trapped in the vortex, a significant amount of dust remains once the dust trapping in the vortex shuts off pebble accretion. This results in an inversion of the grain size distribution between the millimetre to centimetre-sized pebbles, with most mass in the 1 mm population. Clearly visible is the point at which the vortex separates from the planet, marked by the sudden transition to a period of constant mass in the centimetre grain population, until the torque between the planet and vortex during a horseshoe turn leads to a deposition of mass in the planet's Hill sphere, and thus a short period of further accretion. As expected, there is little change in mass of the micron-sized dust population, as their accretion rate is negligible. The total dust mass remaining in the disc is $\approx 50 M_\oplus$, which is trapped in the new pressure maxima that form either side of the planet's orbit, and in the co-orbital region.

\begin{figure*}
  \centering
  \includegraphics{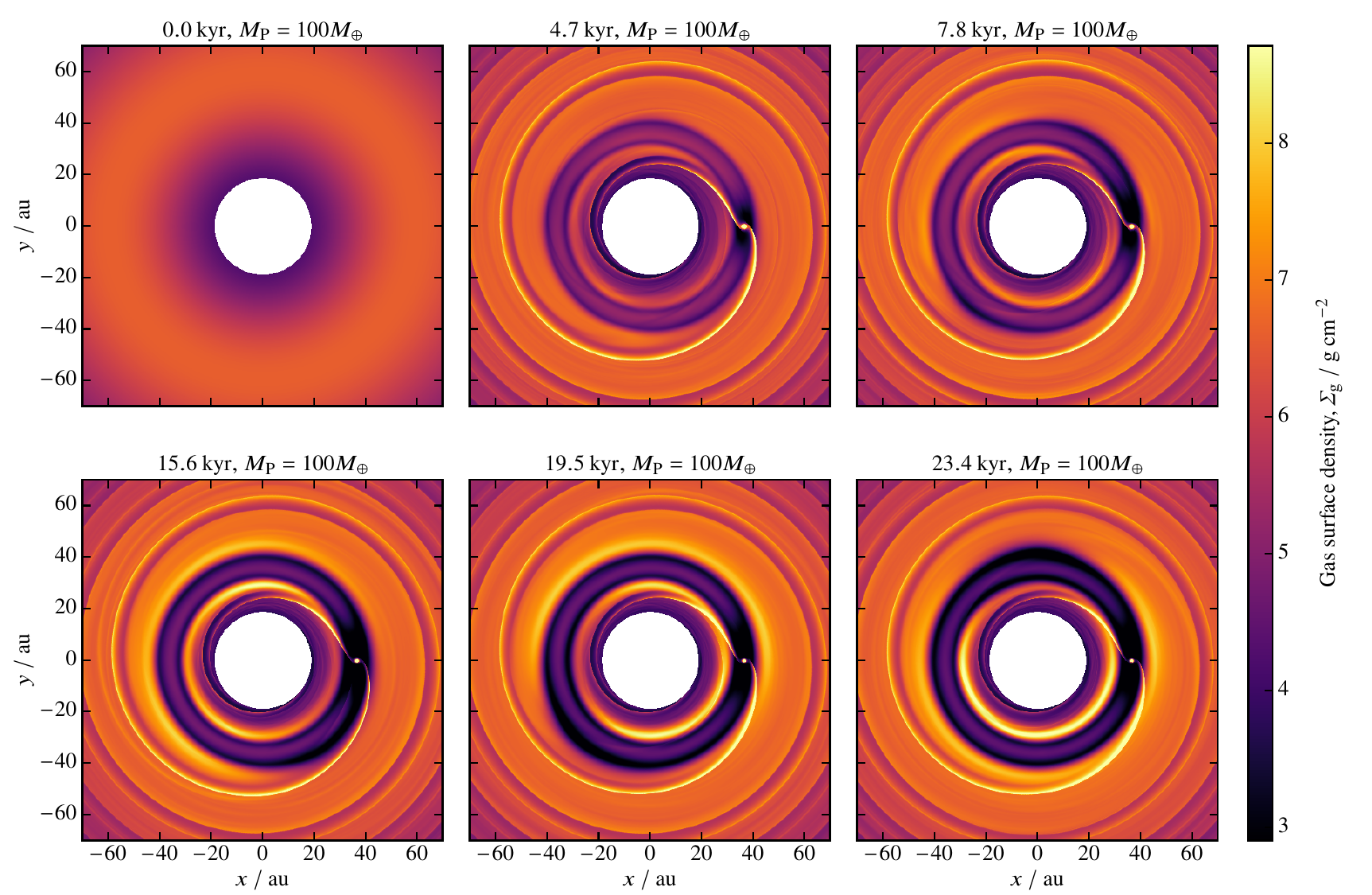}
  \caption{Gas surface density snapshots for the simulation in which the planet is initialised at its final mass and no accretion takes place, thus there is no thermal feedback on the disc. The 100$M_\oplus$ planet carves a deep gap in the disc, the edges of which quickly become RWI-unstable, leading to an azimuthal asymmetry either side of the gap.}
  \label{fig:surfdens_na}
\end{figure*}

\subsubsection{High planet mass test}
Given the results we obtained in the previous simulations, we performed an additional simulation with the same initial conditions but the planet instead initialised with a mass of $100 M_\oplus$. This planet is not undergoing pebble accretion, to further discern which behaviour results solely from the accretion luminosity of the planet rather than through gravitational planet-disc interactions. Figure~\ref{fig:surfdens_na} shows snapshots of the gas surface density from this simulation. The planet interacts gravitationally with the disc and quickly creates a deep gap, the edges of which become RWI-unstable, leading to the formation of a vortex either side of the gap within 10 kyr. The vortices that form are weak, being narrow in radius and wide in azimuthal extent. As was the case in \textsection\ref{sec:mainresult}, the micron and millimetre sized grains also form azimuthally asymmetric substructure due to the vortices, circulating with the gas, but the centimetre grains, being less tightly coupled to the gas, largely maintain a ring structure due to the vortex not being strong enough to trap the dust. Again, dust in the co-orbital region accumulates in the drag-modified L$_5$ Lagrange point, but not in L$_4$.

\subsubsection{Resolution test}
\label{sect:resstudy}
\begin{figure}
  \centering
  \includegraphics{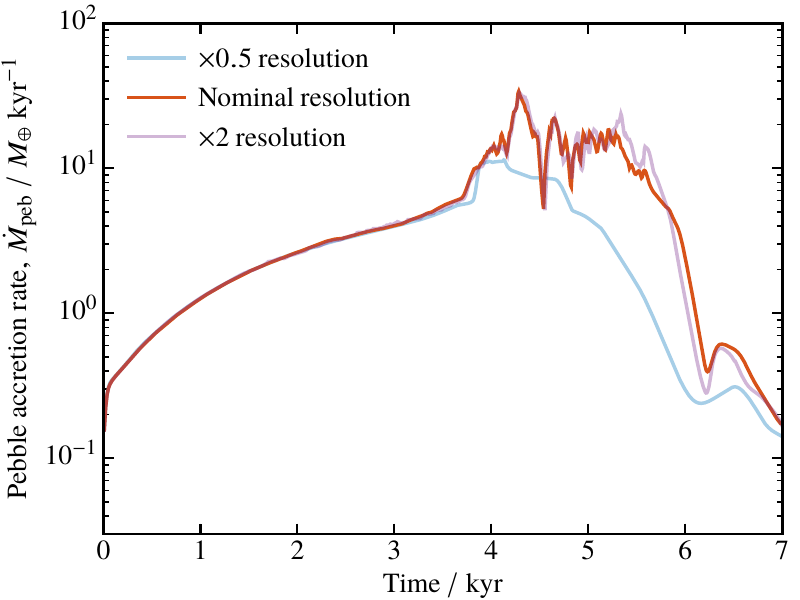}
  \caption{Comparison of the pebble accretion rate onto the planetary embryo as a function of time for half and double the original resolution over the first 7 kyr of the simulation. Pebble accretion from the disc is resolved at half the original resolution, but once the vortex forms at around 3.5 kyr, the lowest resolution case cannot accurately resolve the density gradients around the embryo, reducing dust flow into the Hill sphere and therefore impacting the pebble accretion rate.}
  \label{fig:par_rescomparison}
\end{figure}

We carried out simulations at various resolutions to check that the pebble accretion onto the planet had converged, i.e. that our pebble accretion algorithm was not impacted by the grid resolution. Figure~\ref{fig:par_rescomparison} compares the pebble accretion rate of centimetre-sized dust grains onto the planetary embryo over the first 7 kyr of the simulation -- which includes the initial embryo growth and vortex formation -- at half and double the original resolution. The accretion onto the embryo from the disc is resolved at half resolution, but once the vortex forms at around 4 kyr, the accumulation of dust in the vortex cannot be sufficiently resolved: concentration of dust in the pressure maximum at the centre of the vortex is impeded only by diffusion, and with a low viscosity of $\alpha = 10^{-4}$, the half resolution case cannot resolve the pressure and density gradients upon which the dust dynamics depend. Since the vortex forms at the planet location and the pebble accretion rate depends on the surface density of solids within its Hill sphere, the unresolved accumulation of dust in the vortex restricts the accretion rate onto the planet.

The pebble accretion rate at double the original resolution is qualitatively similar during vortex formation. Since differences in the accretion rate lead to differences in the planet mass, which in turn determines the subsequent accretion rate, small differences are likely to be amplified. However, given the absolute differences in the accretion rate are small, we consider the simulation at our fiducial resolution to be a reasonable compromise between convergence and computing time.

\subsection{Radiative transfer simulations}
In order to understand how the dust distribution calculated from our hydrodynamics simulation might present in observations, we performed Monte Carlo radiative transfer simulations using \texttt{RADMC-3D} \citep{radmc3d}. The dust opacities were calculated using the model presented in \cite{dsharp5}: the dust grains are taken to be of a mixture of ``astronomical silicates'' \citep{draine03}, troilite and ``organics'' \citep{henningstognienko96} and water ice \citep{warrenbrandt08}, with a water ice mass fraction of 0.2. The grains are taken to have zero porosity. The optical constants for this mixture are obtained from their constituents' via the Bruggeman mixing rule, which are then used to calculate the absorption and scattering opacities as a function of wavelength via Draine's implementation\footnote{\url{https://www.astro.princeton.edu/~draine/scattering.html}} of the \texttt{BHMIE} code \citep{bohrenhuffman98}.

We only consider the region of the disc between 22 and 52 au, since this contains essentially all of the millimetre to centimetre sized dust. This region is initially covered by 970 cells in the radial direction, but we reduce the resolution over a few narrow annuli where the dust surface density is approximately constant; this reduces number of radial cells to 435. The azimuthal resolution is reduced by a factor of 6, to 1280 cells. We then perform 2D cubic spline interpolation over the surface density onto this reduced resolution grid. We use 145 cells in the polar direction, with 72 logarithmically spaced between $10^{-3}$ and $\uppi/6$ both above and below the midplane at $\theta = \uppi/2$.

Since the gas pressure structure of the disc prevents the radial drift of dust, and the millimetre and centimetre sized grains have short vertical settling time-scales, we may assume that each dust species has reached equilibrium between diffusion and settling in the vertical direction. We therefore calculate their three-dimensional density distributions as \citep[e.g.][]{takeuchi02}
\begin{equation}
  \label{eqn:3ddust}
  \rho_\mathrm{d} = \rho_\mathrm{d, mid} \exp \bigg[-\frac{z^2}{2H_\mathrm{g}^2} - \frac{\mathrm{Sc}\,\mathrm{St_{mid}}}{\alpha}\bigg(\mathrm{exp}\frac{z^2}{2H_\mathrm{g}^2}-1\bigg) \bigg].
\end{equation}
 The dust density at the disc midplane is calculated via
\begin{equation}
  \rho_\mathrm{d, mid} = \mathit{\Sigma}_\mathrm{d} \bigg\{ \int_{-\infty}^{\infty} \exp \bigg[-\frac{z^2}{2H_\mathrm{g}^2} - \frac{\mathrm{Sc}\,\mathrm{St_{mid}}}{\alpha}\bigg(\mathrm{exp}\frac{z^2}{2H_\mathrm{g}^2}-1\bigg) \bigg] \mathrm{d}z \bigg\}^{-1}
\end{equation}
which we evaluate numerically. Again we set $\mathrm{Sc} = 1$, i.e. we are assuming that the vertical and radial diffusion coefficients are equal. 

The properties of the central star are chosen to be those of a Herbig Ae/Be star. We use an effective stellar temperature of $9000\:\mathrm{K}$, a mass of $2 M_\odot$ and radius of $2.5 R_\odot$. We use a template spectrum for an A0V star from \cite{castelli_kurucz03}, with a wavelength range spanning $1\:\upmu\mathrm{m}$ to $10\:\mathrm{cm}$ and sample this spectrum using $2 \times 10^{10}$ photon packets. The disc is taken to be at a distance of 140 pc, and the resulting image is convolved with a Gaussian beam of FWHM 40 mas ($\approx 5$ au at 140 pc), such that the resolution is comparable to that achieved in the DSHARP survey. We take a representative noise floor from the DSHARP survey of 15 $\upmu\mathrm{Jy\:beam^{-1}}$ and remove any signal below this value from the images. We generate images at observation wavelengths of 0.88 mm and 1.25 mm for comparison with ALMA band 7 and 6 observations respectively.

\begin{figure*}
  \centering
  \includegraphics{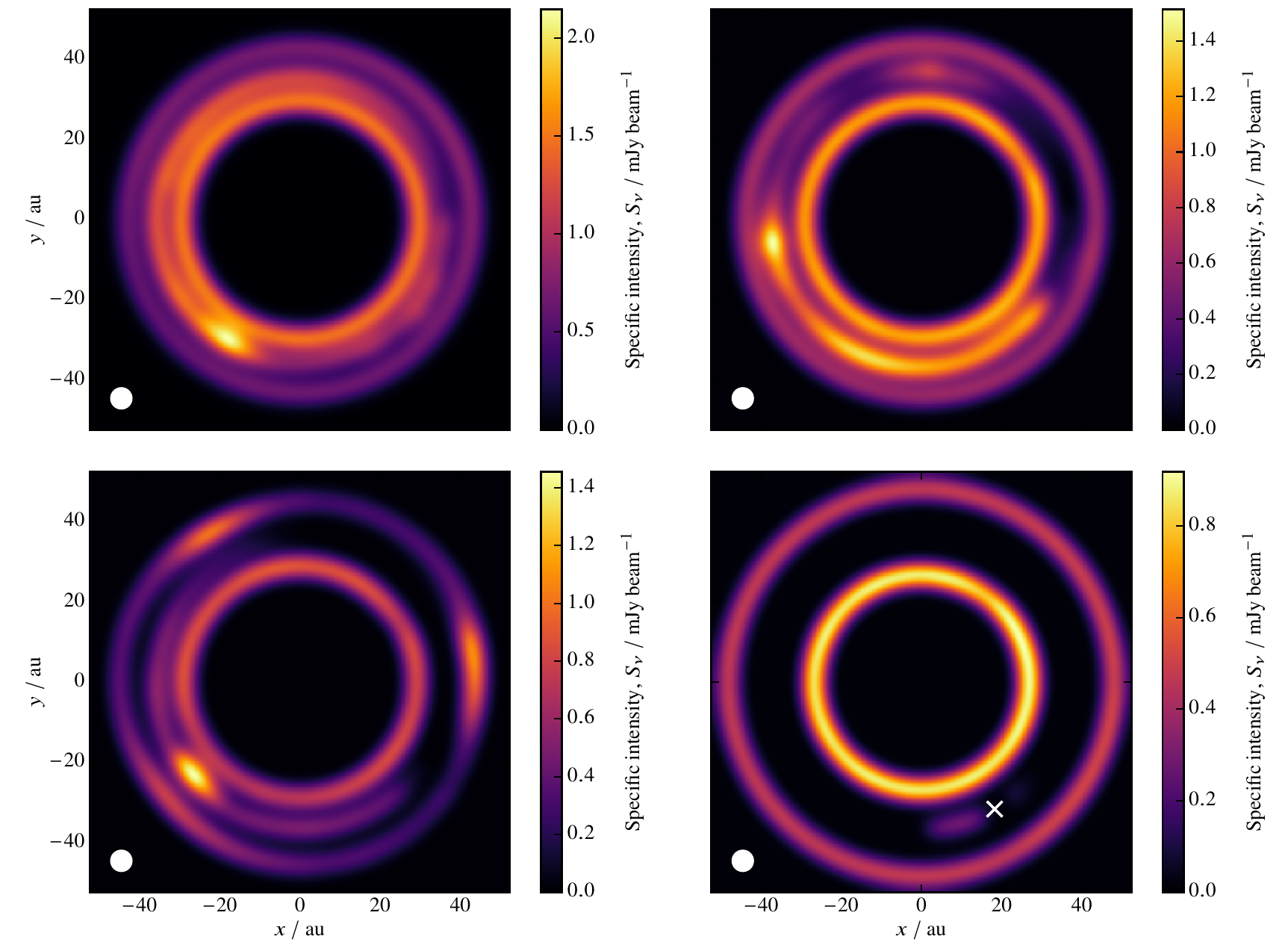}
  \caption{Synthetic observations of the disc at 11.7, 14.1, 22 and 226 kyr, using an observation wavelength of 1.25 mm, convolved with a 5 au FWHM Gaussian beam (shown in the bottom-left of each panel). Flux below our adopted instrumental noise floor of 15 $\upmu\mathrm{Jy\:beam^{-1}}$ has been removed. In each image a brightness enhancement can be associated with a dust surface density enhancement due to trapping in vortices. In the final image, the dust residing close to L$_5$ (marked with a white cross) can be seen, with less flux from the largest dust to the right of L$_5$ compared to the millimetre sized dust, which has settled to the left of L$_5$.}
  \label{fig:synthobs}
\end{figure*}

Figure~\ref{fig:synthobs} shows the images generated at an observation wavelength of 1.25 mm from simulation snapshots at 11.7 kyr, 14.1 kyr, 22 kyr and 226 kyr. In the early images, while dust is still present in the co-orbital region, the brightness enhancements resulting from dust trapped in vortices are resolved, as is the lack of emission from the vicinity of the planet. At 22 kyr, there is significant emission from the dust that has been dragged away from L$_5$ by the vortex, while the compact vortex structure stands out. Also visible is the $m=3$ vortex pattern in the outer ring, resulting from the RWI. 

\begin{figure*}
  \centering
  \includegraphics{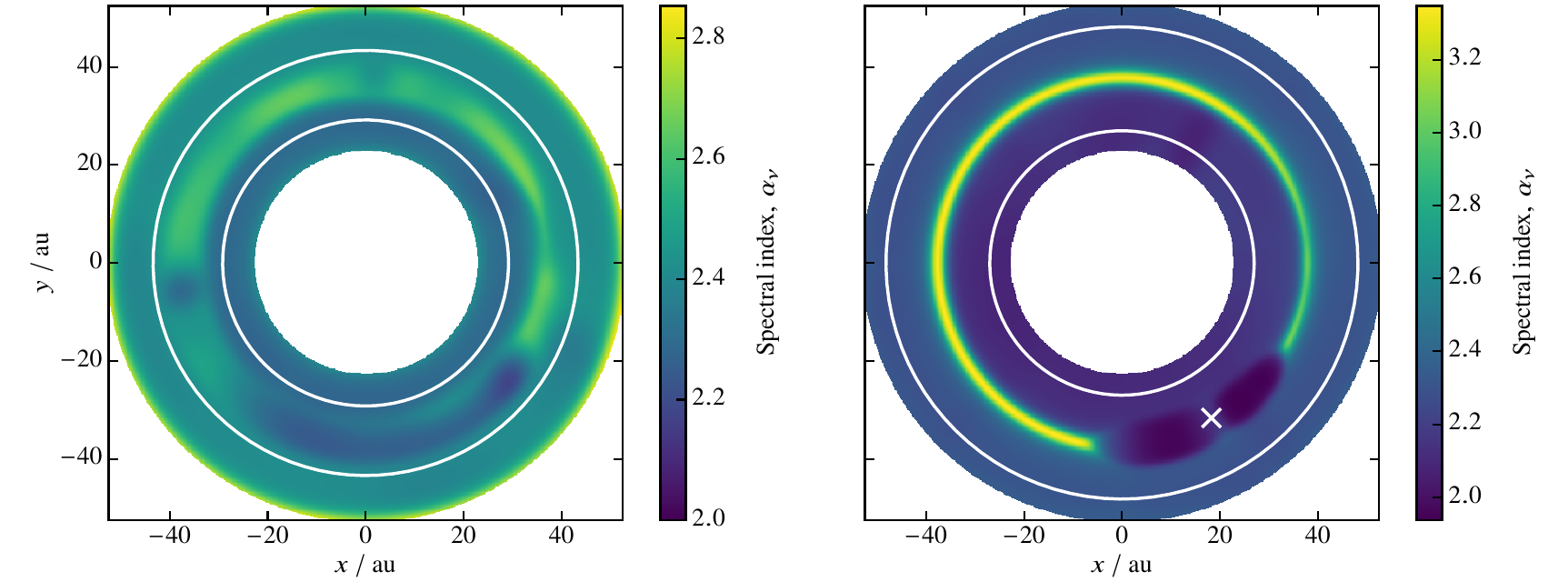}
  \caption{Spectral index map of the disc at 14.1 kyr (left) and 226 kyr (right), calculated between 0.88 mm and 1.25 mm. At 14.1 kyr, the low spectral index regions at around (2, 35) and (-35, -5) are due to dust trapping in vortices, while the crescent spanning (-20, -30) to (30, -20) is due to dust beginning to settle around L$_5$. At 226 kyr, an extended region of low spectral index can be seen around L$_5$, which is marked with a white cross. The white solid lines mark the locations of the pressure maxima.}
  \label{fig:specidx}
\end{figure*}

The general morphology resulting from the band 6 and 7 synthetic observations is similar, though we do see flux ratio variations. We calculated the spectral index, $\alpha_\nu = \mathrm{d}\log I_\nu/\mathrm{d}\log\nu$, over the disc between these two wavelengths. The images are each convolved with the Gaussian beam before the spectral index calculation. As shown in figure~\ref{fig:specidx} the spectral index is lowest in the dust rings and at the vortex centres, indicating the concentration of the largest dust grains there. Similarly, the spectral index is low around L$_5$, where dust grains are trapped. Due to the fact that different sized dust grains settle at different locations around L$_5$, the region of low spectral index has a much wider azimuthal extent than the vortex, which essentially just presents as a single beam-sized region of low spectral index. Note that the separation between the two regions of emission around L$_5$ in the final image arises due to our use of discrete grain sizes in our hydrodynamics simulations; a continuum of grain sizes would produce a continuous crescent of emission.

\section{Discussion}\label{sec:discuss}
\subsection{Disc morphology}
We set out to explore a new link between planet formation and non-axisymmetric substructure in protoplanetary discs. Our simulations have shown that a planetary embryo in an axisymmetric dust trap can accrete pebbles at a high enough rate to trigger the formation of vortices and thus create non-axisymmetric substructure in the disc. The vortex that forms spans $\approx 15-20$ au in azimuthal extent at 35 au from the central star. Since dust drifts towards the pressure maximum located at the centre of the vortex, the size of the non-axisymmetric enhancement in dust will vary over time as well as with grain size. As the strength of the vortex decays it releases dust and its size therefore decreases over time, while dust drift, limited only by diffusion and being dependent on grain size, sets the observable size of the dust asymmetry. Our synthetic observations show that the substructure generated in our simulation creates detectable features such as rings and crescents similar to those observed in continuum images \citep[e.g.][]{dsharp1}, as well as features in spectral index maps \citep[e.g.][]{huang18, carrascogonzalez19, huang20}.

Having accreted a significant fraction of the mass from the dust trap, the planet becomes massive enough to carve a gap in the initial pressure maximum, creating new pressure maxima either side of its orbit. Dust becomes trapped in these two pressure maxima, leading to the presence of two dust rings which exist alongside the vortex that forms. The dust remaining in the disc, which primarily resides in the pressure maxima either side of the planet's orbit, shows an inversion of the pebble size distribution, with most mass in 1 mm grains and the least in the 1 cm grain population. Given time, however, we suspect an MRN distribution could be re-established through collisions via coagulation and fragmentation.

\subsection{Properties of the planet and a new phase of pebble accretion}
In addition to the non-axisymmetric disc substructure produced through this mechanism, we find interesting consequences for the growth of the planet. Our simulation produces a $100 M_\oplus$ core, while in our test case where we do not include the accretion luminosity, the planet grows to $\approx 40 M_\oplus$. This is naively opposite to what one might have imagined, where the vortices trap dust and migrate away from the planet, preventing this dust from accreting onto the planet, resulting in a lower planet mass. However, since the vortex forms around the planet and remains co-located with it for $\sim 30$ orbits, it acts to significantly increase the area from which the planet can accrete before migrating away and trapping dust.

\begin{figure}
  \centering
  \input{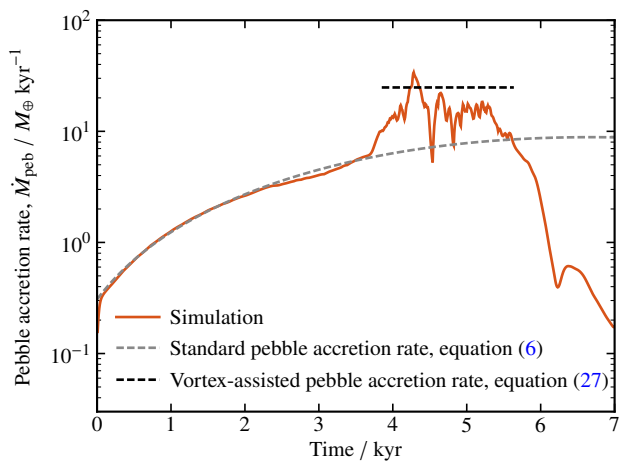}
  \caption{Comparison of the pebble accretion rate in the simulation including thermal feedback with the theoretical pebble accretion rate according to the standard pebble accretion scenario, as well as the theoretical accretion rate during the vortex-assisted pebble accretion phase, time-averaged over the period when the vortex is co-located with the planet.}
  \label{fig:mpebdot_comparison}
\end{figure}

As the vortex is a pressure maximum, it acts to funnel dust towards the planet from a radial width of $\sim 4/3H_\mathrm{g}$, rather than $\sim 2R_\mathrm{H}$ in the case of standard pebble accretion. Thus, our simulations have identified a new, short-lived, but extreme form of vortex-mediated pebble accretion. It is the much larger feeding zone that allows the planetary core to reach such a large mass. In this phase the ``vortex-assisted'' pebble accretion rate,
\begin{equation}
  \dot{M}_\mathrm{peb, vort} \approx \frac{4}{3}\mathit{\Omega}_\mathrm{K} H_\mathrm{g}^2 \mathit{\Sigma}_\mathrm{d},
  \label{eqn:m_peb_dot_va}
\end{equation}
is actually independent of the planet mass, and hence time. This explains the sudden increase, but approximately constant (when averaged over several orbital periods) pebble accretion rate seen in our simulations. Taking our planet mass as $\sim 15 M_\oplus$ at 4 kyr (figure~\ref{fig:planetmass}), the change in effective accretion area from $2R_\mathrm{H}^2$ to $4/3 H_\mathrm{g}^2$ should result in a jump in accretion rate of $\sim 4$. 
This is illustrated in figure~\ref{fig:mpebdot_comparison}, where the theoretical classical pebble accretion rate and this theoretical vortex-assisted pebble accretion rate are plotted. The classical pebble accretion rate is obtained by numerically integrating equation~(\ref{eqn:m_peb_dot}) while accounting for the depletion in pebbles, and the theoretical vortex-assisted accretion rate is obtained from equation~(\ref{eqn:m_peb_dot_va}), using the time-averaged dust surface density over the period when the vortex is co-located with the planet. This is approximately in agreement with the jump in accretion rate from $\sim 6 M_\oplus$ kyr$^{-1}$ at 4 kyr to a time averaged rate of $\sim 20 M_\oplus$ kyr$^{-1}$ seen during the phase where the vortex is co-located with the planet between 3.5 and 6 kyr. Thus, we suspect this new form of vortex-assisted pebble accretion could play an important role in planet formation inside dust traps, at the extreme end of planet formation.

Our simulations, both with and without thermal feedback, result in higher masses than typically found in previous models. This is because our planets form inside a dust trap where there is already a large supply of dust. Standard models of pebble accretion, which typically use a disc with a monotonically decreasing surface density profile, find a maximum core mass that can be formed: the pebble isolation mass. This is the mass at which the core can prevent further drift of pebbles into its orbit; once it is massive enough to carve a gap, the resulting pressure bump exterior to its orbit prevents further inward drift of pebbles \citep{morbidelli12, lambrechts14, bitsch18}. It is worth noting that the scenario in which this standard pebble isolation mass arises is not applicable here, since in a dust trap the planet accretes pebbles within the proximity of its own orbit rather than pebbles continuously drifting in from the outer disc, so halting pebble drift from outer disc has no effect on planet, hence why it can grow so large.

This was also recently demonstrated by \cite{sandor_regaly21}; they performed 2D hydrodynamics simulations of a core undergoing pebble accretion in a pressure maximum, using the \cite{kley99} prescription for gas accretion as their pebble accretion scheme. Neglecting accretion luminosity, they found a rapid pebble accretion rate and core growth up to final masses between $\sim 30 - 50M_\oplus$, depending on their prescribed pebble accretion efficiency and the Stokes number used. In their control case without a pressure maximum, cores reached $\approx 20M_\oplus$, in agreement with \cite{lambrechts14}.

While these solid masses seem large, there is strong evidence for large solid masses in the observed exoplanet population. Thermal and structural modelling of transiting giant planets by \cite{thorngren16} suggests metal masses $M_\mathrm{Z} \gtrsim 100 M_\oplus$ for $\approx 20\%$ (9/47) of their sample of Jupiter mass planets. Assuming that these planets formed via core accretion, the authors suggest that these may be giant planets with metal-enriched atmospheres, though their modelling is not constraining enough to indicate where the solids are partitioned within the planet. Thus, formation of a massive planetary core in a dust trap, followed by subsequent gas accretion (our solid masses are well above the critical core mass for runaway accretion, once the pebble accretion has subsided, e.g. \citealt{Rafikov2006}) and migration is a viable formation scenario of the observed metal enriched giant planets.

\subsection{Final trapping of dust at Lagrange points}
Our simulations, both with and without thermal feedback, show that once the vortex has dissipated, the dust it had accumulated gathers around the classical L$_5$ Lagrange point. Dust trapping at L$_5$ has been seen in previous work: for example, \cite{rodenkirch21} showed that dust trapped in L$_5$ may be responsible for non-axisymmetric feature in observed in HD 163296, which shows a crescent-shaped region of emission in the gap between a dust ring and the inner disc. Their simulations of planets embedded in a disc resulted in a crescent-shaped feature between two rings, similar to the final state of our simulation, thus an asymmetry in the gap between two rings resulting from trapping in L$_5$ is not unique to our model.

The reason why dust is trapped in L$_5$ but not in L$_4$ has been studied analytically within the restricted three-body problem. \citet{murray94} showed that the addition of gas drag modifies L$_4$, making it an unstable equilibrium point. This is confirmed in our simulations by analysing the forces on dust grains in the vicinity of L$_4$. Additionally, \citet{murray94} showed that gas drag modifies the location of the stable L$_5$ point, where the position of L$_5$ relative to the planet depends on the stopping time of the dust grains, with less strongly coupled dust residing closer to the planet. Our simulations generally confirm this trend; however, we do see some deviations from the predictions. This discrepancy is attributed to the fact that \citet{murray94} did not use an exact gas streamline structure, but instead adopted a simple model. 

This dust-size-dependent position of the drag-modified L$_5$ location means that dust trapped in L$_5$ and a vortex have a potential observational difference. In the case of the vortex, the peak dust density is always achieved in the same place -- the vortex core -- irrespective of particle size, resulting in a strongly peaked dust density and therefore a single location where the spectral index drops. In the case of L$_5$, however, the particle distribution is spread over a much wider range in azimuth, resulting in a broad region where the spectral index drops, with a gradient such that the spectral index decreases towards the planet (see figure~\ref{fig:specidx}). This is a potentially important avenue in observationally distinguishing the nature of the observed non-asymmetric structure.

\subsection{Impact of our simplifications}
Given that we wished to isolate the effects of thermal feedback, there are some physical processes that we have neglected from our investigation, the possible implications of which we now discuss. The first is that we have ignored the back-reaction on the gas from the drag force on the dust. The peak dust-to-gas ratio $\mathit{\Sigma}_\mathrm{d}/\mathit{\Sigma}_\mathrm{g} \approx 0.8$ in our initial condition, thus the back-reaction on the gas will be non-negligible here. The effect of this would be to widen the gas pressure maximum and thus weaken the dust trap, which would reduce the dust surface density and hence the initial pebble accretion rate onto the planetary embryo. The feedback on the gas will also become important in the vortices where large amounts of dust become trapped. \cite{fu14} found that the back-reaction can reduce the lifetime of the vortex, and a similar result was found by \cite{raettig15}. However, follow-up work by \cite{lyra18} showed that in 3D simulations the vortex lifetime is unaffected by their accumulation of dust. Regardless of whether the feedback on the gas affects vortex lifetime, it is likely to impact the embryo's growth as described above, and this \textit{does} impact vortex formation, thus this is something we wish to address in future work. As the simulations we have presented here are 2D, it will be necessary to perform 3D simulations to verify our results regardless, once the back-reaction is taken into account.

Since the dust surface density becomes so large in the vortices, it may also be necessary to include self-gravity and see whether it is possible to achieve gravitational collapse of the solids inside a vortex. For example, \citet{lyra09} presented surface densities capable of achieving gravitational collapse inside vortices. Additionally, \citet{lin2018} showed that self-gravity can modify vorticity. While the possibility of gravitational collapse within the vortex is of secondary interest to our investigation, the effect of self-gravity on the lifetime of vortices generated in this way is important to understand.

It is also important to consider the manner in which we remove dust from the embryo's Hill sphere. Due to our treatment of the dust as a fluid, we cannot remove individual dust grains, but instead remove an amount of dust from each cell within the Hill sphere in proportion to the mass currently in that cell. While we have established that our pebble accretion algorithm provides the correct accretion rate, the dynamics of the dust fluid parcels inside the Hill sphere are not explicitly taken into account. We have demonstrated that this is not a problem for standard pebble accretion, but this simplification might have consequences in a situation where the velocity structure differs from that of a standard star-planet system, such as in the presence of a vortex. For example, if a vortex approaches or passes the planet and any dust trapped within the closed streamlines of the vortex enters the planet's Hill sphere, the planet will accrete some fraction of that dust, which may not necessarily represent a physically realistic scenario given the dust dynamics in the vortex. Essentially, our algorithm cannot capture the competition between pebble accretion onto the embryo and drift towards the centre of the vortex. However, a vortex passing through the embryo's Hill sphere is a rare occurrence since they are typically seen to traverse horseshoe orbits in our simulation, so this simplification should have little overall impact. Indeed, the torque between the planet and vortex causes dust to enter the planet's Hill sphere, but the fate of this dust is determined by the equation of motion -- we do not force it all to be accreted by virtue of it entering the planet's Hill sphere. Furthermore, we can be confident that this simplification does not invalidate our results during the initial vortex formation phase, where the vortex forms at the planet location, since figure~\ref{fig:ffcomparison} shows that the pebble accretion rate during this phase is set by how the embryo regulates the mass in its Hill sphere.

An additional simplification we have made in our simulation is the way in which we set the gas temperature in the vicinity of the accreting planet. The temperature profile instantaneously adjusts to the accretion luminosity, which in turn is instantaneously set by the accretion rate. While the outer regions of the Hill sphere and the disc in the simulation are optically thin, the region where the thermal energy resulting from accretion is deposited, deep in the protoplanet's envelope, is likely to be optically thick, resulting in some time-lag for changes in the accretion rates to be translated into temperature changes. For example, in our simulations, once accretion shuts off, the vicinity of the disc around the planet instantaneously cools and the gas surrounding it contracts. While this effect is physical, it is unlikely to be so extreme in a real disc due to the finite time taken for the envelope to cool. In future, simulations should include an improved calculation of the thermal structure around the planet.

Another potential issue that should be investigated is the thermal ablation of pebbles. At the highest luminosities the temperature structure in the envelope, near the core, may exceed the sublimation temperature of the pebbles. If this is the case, (i) the pebbles may not reach the core and (ii) may release their thermal energy at a higher location in the planet's potential; both effects could reduce the accretion luminosity. However, we note that this effect would not prevent the pebbles from actually accreting onto the planet in the first place. 

Another effect that we have neglected from our simulations is the migration of the planet, caused by the exchange of its angular momentum with the disc \citep[e.g.][]{goldreich97}. The type I migration rate of planets with masses less than a few tens of an Earth mass scales linearly with planet mass \citep[e.g.][]{ward97}. However, these migration time-scales are still longer than the planet growth time-scales we find in our simulations. 

\cite{guilera_sandor17} performed 1D simulations of a disc containing axisymmetric pressure maxima and a planetary embryo of 0.01$M_\oplus$, free to migrate as it grew via pebble accretion. The difference between their calculated pressure maxima and zero-torque locations were minimal; as such they found that embryos which formed beyond 2 au migrated towards the pressure maximum, at which point pebble accretion ensued; at even further separations the planet's migration was limited ($\Delta a \lesssim 1$ au) and pebble accretion was still rapid without the planet reaching the pressure maximum. In each case their core mass exceeded $50M_\oplus$ before the simulations were terminated.

In addition, \cite{morbidelli20} investigated the growth of a planet by pebble accretion in a dust ring with properties motivated by the DSHARP observations. Following the growth of an initially $0.1 M_\oplus$ planetary embryo over 3 Myr, the torques on the planet were calculated using the \cite{paardekooper11} formalism, which resulted in minimal migration of the planet, such that the planet remained close to the pressure maximum. As such, its growth was approximately the same as in the case without migration, with no difference in the final planet mass. 

Recent investigations have revealed further possibilities for planet migration. For example, \cite{benitezllambay18} showed that asymmetries in the dust distribution close to the planet may exert a net torque on the planet, the strength of which increases with the dust mass, and pebble accretion itself may compound this \citep{regaly20}. Heating from the planet can also give rise to an additional torque contribution, known as the ``thermal torque'' \citep[e.g.][]{Masset2017,fromenteau19}, which has been shown to modify the migration, including generating outward migration and potentially changing the outcome of pebble accretion \citep[e.g.][]{baumann20,guilera21}. Thus, planetary migration should be explored in future work to see how it might impact both vortex formation and planetary growth. 

\subsection{Comparison to previous work}
As we have built on the work of \cite{owen17}, it is pertinent to compare our results to theirs and thus assess the importance of self-consistently calculating the pebble accretion rate, in particular with respect to the impact on the planet mass achieved and on vortensity generation. In their simulations, \cite{owen17} held the planet mass at a constant $M_\mathrm{P} = 1.5 \times 10^{-5} M_*$, a factor of 100 larger than our initial embryo mass. This difference in initial embryo mass is inconsequential since the initial growth of the embryo in our simulation is rapid, reaching $1.5 \times 10^{-5} M_*$ after 20 orbits, which is before vortex formation begins to affect the pebble accretion rate. What does matter is the subsequent growth of the planet, since it reaches $\sim 80 M_\oplus$ before the vortex migrates from the planet location, which is large enough to carve a gap in the disc. This is in part driven by this new, extreme phase of vortex-assisted pebble accretion our simulations have identified. Gap formation does not occur in \cite{owen17} since their fixed-mass planet is simply not massive enough to strongly perturb the disc, and the extreme vortex-assisted pebble accretion rates were not anticipated. 

The use of a constant planet mass meant that their accretion luminosity was not coupled to the accretion rate, but was instead increased linearly over time until 50 planetary orbits, after which it was kept constant. This provided a sustained source of vortensity which allowed multiple vortices to form and ultimately merge into a single, large-scale vortex. In contrast, since our accretion rate is coupled to the local dust surface density, the accretion luminosity quickly drops off once the vortex carries dust away from the planet, which prevents further vortex formation. 

In \cite{owen17} it was envisioned that once the vortex dissipates and releases the dust it had accumulated, the ring would re-form, restarting accretion onto the planet and potentially the vortex formation process. However, we find that the accretion rate required to initially form a vortex yields a planet massive enough to carve a gap in the disc, which prevents dust from entering its Hill sphere if the ring were to re-form. Moreover, our simulations have shown that dust released from the vortex settles in the (drag-modified) L$_5$ Lagrange point, and is only removed through diffusion, which is a slow process given expected viscosity values. Dust released from the dissipating vortex would therefore not immediately form a ring, and any further accretion of this material onto the planet would be at a rate insufficient to trigger vortex formation.

Since our initial simulations have been performed at the extreme end of the parameter space it is still unclear whether the cycle anticipated by \citet{owen17} can occur or not. For example, if at lower initial dust masses a vortex can form but with a lower planet mass, dust will be less strongly trapped at L$_5$, and may also be able to accrete onto the planet once the vortex has dissipated because of the planet's reduced gap-carving potential.

\section{Conclusions}
We have investigated the growth of a pebble-accreting planetary embryo in an initially axisymmetric dust trap within a protoplanetary disc. We performed global hydrodynamics simulations of the dust and gas at a resolution high enough to resolve the pebble accretion process, allowing us to couple the accretion rate onto the embryo to the local dust surface density. By considering the accretion luminosity of the planetary embryo, we have shown that if the embryo can heat the disc beyond its Hill sphere, the gas becomes locally unstable to the formation of anticyclonic vortices. We found that the formation of a vortex at the location of the planet has significant consequences for the planet's growth: since an anticyclonic vortex is a local pressure maximum, dust drifts towards the vortex and therefore towards the planet. The vortex thus provides a larger area for dust supply into the embryo's Hill sphere than permitted by the planetary potential alone. This ``vortex-assisted'' pebble accretion ultimately leads to a higher final planet mass than can be achieved without considering the embryo's accretion luminosity.

The planetary embryo accretes $\sim 100 M_\oplus$ of the $150 M_\oplus$ in the dust trap, resulting in a large amount of solids in the planet, perhaps providing a formation mechanism for giant planets that are inferred to be extremely metal rich \citep[e.g.][]{thorngren16}. The primary reason for the growth of such a large core is, as discussed above, the formation of the vortex at the planet location. When not accounting for the embryo's accretion luminosity however, the embryo grows to $\sim 40 M_\oplus$, which is still relatively large for a planetary core. The reason for this large core growth is due to the high surface density of pebbles in the dust trap, which is a result of the high disc dust mass and the viscosity assumed. While our disc parameters are observationally motivated, as non-axisymmetric discs are typically around Herbig Ae/Be stars which tend to have more massive discs, it is important to understand the range of parameter space in which this vortex formation process is able to occur. Thus, this provides motivation for a parameter study in future work.

Our simulations have also demonstrated a new link between planet formation and non-axisymmetric substructure in protoplanetary discs. While this mechanism does not produce vortices on the scale required to explain the most extreme asymmetries such as those observed in IRS 48 and HD 142527 -- at least for the disc parameters we have used here -- it may have some promise in explaining the non-axisymmetric features observed in HD 163296 and HD 143006, as well as those observed in recent DM Tau observations. The vortex survives for $\sim 75$ kyr ($\sim 500$ orbits), though the dust asymmetry lasts for $\sim 50$ kyr -- just a fraction of the typical $\sim$ Myr lifetime of protoplanetary discs. How the size and lifetime of the vortex and observable dust asymmetry varies with initial dust mass and physical disc properties such as temperature and viscosity will be the focus of our upcoming parameter study, in order to better understand the potential of this mechanism.

\section*{Acknowledgements}
We thank the anonymous referee, whose review helped improve the clarity of this paper.
DPC thanks Matth\"{a}us Schulik for helpful discussions.
DPC is supported by a 2017 and 2020 Royal Society Enhancement Award. JEO and RAB are supported by a Royal Society University Research Fellowship. This project has received funding from the European Research Council (ERC) under the European Union’s Horizon 2020 research and innovation programme (Grant agreement No. 853022, PEVAP). 
This work was performed using the DiRAC Data Intensive service at Leicester, operated by the University of Leicester IT Services, which forms part of the STFC DiRAC HPC Facility (www.dirac.ac.uk). The equipment was funded by BEIS capital funding via STFC capital grants ST/K000373/1 and ST/R002363/1 and STFC DiRAC Operations grant ST/R001014/1. DiRAC is part of the National e-Infrastructure.
This work was performed using the Cambridge Service for Data Driven Discovery (CSD3), part of which is operated by the University of Cambridge Research Computing on behalf of the STFC DiRAC HPC Facility (www.dirac.ac.uk). The DiRAC component of CSD3 was funded by BEIS capital funding via STFC capital grants ST/P002307/1 and ST/R002452/1 and STFC operations grant ST/R00689X/1. DiRAC is part of the National e-Infrastructure.
For the purpose of open access, the authors have applied a Creative Commons Attribution (CC-BY) licence to any Author Accepted Manuscript version arising.

\section*{Data availability}
The data underlying this article will be shared on reasonable request to the corresponding author. Videos of the simulations presented in this article are publicly available on Zenodo, at \href{https://doi.org/10.5281/zenodo.6761702}{10.5281/zenodo.6761702}.



\bibliographystyle{mnras}
\bibliography{main}





\bsp	
\label{lastpage}
\end{document}